\newcommand{\COMMENT}[1]{}

\newcommand{\reff}[1]{}           

\newcommand{\spaceline}[1]{\vspace{9pt}}
\newcommand{\spacex}[1]{\vspace{1cm}}
\newcommand{\spacexx}[1]{\vspace{2cm}}

\newcommand{\kl}[1]{\left\langle} 
\newcommand{\kr}[1]{\right\rangle} 

\documentclass[10pt,a4paper]{article}  
\usepackage{graphicx}
\usepackage{amssymb}
\usepackage{amsmath}

\usepackage[utf8]{inputenc}  
\usepackage[T1]{fontenc}

\usepackage{here}  

\usepackage[font=small,margin={9mm,9mm}]{caption} 

\usepackage[colorlinks]{hyperref}

\def\DREIECK#1{{\def\bull{}%
\count1=0
\loop
\edef\bull{$\bullet$\bull}
\ifnum\count1<#1
\advance\count1 by 1
\centerline{\bull}
\vskip-7.7pt
\repeat
\vskip 7.7pt\relax}}

\begin{document}
\setlength\parindent{0pt}
\setlength{\parskip}{6pt}  
\setlength{\overfullrule}{10pt}


\begin{center}
\Large
\textbf{Optimizing  Wealth by a Game within Cellular Automata}
~~\\
~~\\
\normalsize
Rolf Hoffmann$^1$,
Franciszek Seredy\'{n}ski$^2$,
Dominique D\'{e}s\'{e}rable$^3$\\
\footnotesize
~~\\
$^1~$Technical University Darmstadt, Germany\\
{hoffmann@informatik.tu-darmstadt.de}\\
~~
$^2~$Institute of Computer Science, University of Siedlce, Poland\\
{franciszek.seredynski@uws.edu.pl}\\
~~
$^3~$Institut National des Sciences Appliqu\'{e}es, Rennes, France\\
{domidese@gmail.com}

\end{center}

\begin{abstract}
\noindent
The objective is to find a Cellular Automata (CA) rule that can evolve 2D patterns 
that are optimal with respect to a global fitness function.
The global fitness is defined as the sum of local computed utilities.
A utility or value function  computes a score depending on 
the states in the local neighborhood. 
First the method is explained that was followed to find such a CA rule. 
Then this method is applied to find a rule that maximizes social wealth. 
Here wealth is defined as the sum of the payoffs that all players
(agents, cells) receive in a prisoner's dilemma game, and then shared equally among them.
The problem is solved in four steps:
(0) Defining the utility function,
(1) Finding optimal master patterns with a Genetic Algorithm,
(2) Extracting templates (local neighborhood configurations),
(3) Inserting the templates in a general CA rule. 
The constructed CA rule finds optimal and near-optimal patterns for 
even and odd grid sizes. 
Optimal patterns of odd size contain exactly one singularity, a 2 x 2 block of 
cooperators. 
\end{abstract}

\normalsize
%

%
\newpage
\tableofcontents
\newpage

\section{Introduction}

The general topic is to form certain patterns of colored cells arranged in a grid/field of discrete points. 
We aim at patterns that are optimal with respect to a global objective/fitness function over the field of cell states/colors.
Specifically, we want to find the global optimum by the use of local operations only. 
Cellular Automata (CA) act locally and effect globally, 
and it is well known that the iterated application of a local rule can form complex patterns. 
Therefore we try to solve our problem with CA.  

The goal is to find a CA rule that can form  2D patterns 
which are optimal with respect to a given  fitness function.
The fitness function shall be the sum of \textit{local value functions} (also called \textit{utilities}) computed by every cell taking into account their neighboring states.
In this work we define the utility as the payoff/reward 
that a cell/agent/player receives out of a game played with his neighbors,
the Prisoner's Dilemma Game. 
Note that only the CA rule determines the actions (cell state changes) of the
``game'' and there are no separate players.
The payoff function of the game is only used as an example for the definition of a
utility function. 

In this article we use the following terms. 
A \textit{cell} is a entity at a certain site in a 2D grid,
compromising a \textit{state}, and an \textit{automaton} (a \textit{rule}) with \textit{connections} 
to its local neighbors. 
A array of cells is termed \textit{Cellular Automata} (CA).
A CA cell is also called \textit{agent}, because it can change states, 
temporary variables, and
 optionally some states of the environment. 
The agents used here are simple cells that change their own state only,
but they can compute in addition the   \textit{utility} function, 
a  value for a certain local neighborhood configuration. 


\textbf{Interpretation of our special problem.}
The pattern we aim at will be a mixture of 
zeroes (cooperators in a game, displayed in white/green/yellow),  
and ones (defectors, displayed in black/blue).
An optimal pattern will contain around 1/4 defectors and 3/4 cooperators
locally arranged as ``points''
($\substack{000
          \\010
          \\000}$)   
and dominoes 
($\substack{0000
          \\0110
          \\0000}$)  
of defectors.
We can interpret the pattern as a community of agents,
defectors and cooperators. 
We can name the defectors ``dealers'' or enterprisers, and the 
cooperators ``consumers'' or workers.
We know from economics that the prosperity or 
\textit{wealth of the whole community} 
increases if there is a certain amount of dealers among all agents. 
We require that the incomes/utilities of all dealers and all consumers
of the community are shared, equally distributed among all of them. 
So we consider a systems where all agents behave fully social. 
The system tries to optimize its wealth as a whole and without giving any
preference to individuals.

\COMMENT{
Nash, J.F., Equilibrium points in n–person games, PNAS USA 36(1) (1950) 48-–49

\bibitem{Rapoport1965}
Rapoport, A., Chammah, PL., Prisoner’s Dilemma, Univ. Michigan Press (1965)

\bibitem{Axelrod1984}
Axelrod, R., The Evolution of Cooperation, Basic Books, Inc., New-York (1984)

\bibitem{Nowak1992}
Nowak, M., May, R., Evolutionary games and spatial chaos, Nature 359 (1992)
826–829

\bibitem{Nowak1992(2)}
Nowak, M.,Sigmund, K., Tit–for–Tat in heterogeneous populations, Nature 355
(1992) 250–253

\bibitem{Nowak1992(3)}
Nowak, M., Five rules for the evolution of cooperation, Science 314 (2006) 1560–1563

\bibitem{Grim1995}
Grim, P., The greater generosity of the Spatialized Prisoner’s Dilemma, J. Theor.
Biol. 173(4) (1995) 353–359

\bibitem{Nagpal2008} 
Nagpal, R., Programmable pattern–formation and scale–independence, Unifying
Themes in Complex Systems IV, Minai, A.A, Bar–Yam, Y., eds., (2008) 275–-282

\bibitem{Yamins2008}
Yamins, D., Nagpal, R., Automated Global–to–Local programming in 1D spatial
multi–agent systems, Proc. 7th Int. Conf. AAMAS (2008) 615–-622
}

\textbf{Related Work.}
The \textit{Prisoner's Dilemma} is very well known and studied in the theory of games 
\cite{
Axelrod1984,
Grim1995,
Nash1950,
Nowak1992,
Nowak1992(2),
Nowak2006,
Rapoport1965}.
The prisoner dilemma's game is used as a mean to optimize systems and patterns
in \cite{
 Katsumata2008,
Tretyakova2014,
Seredynski2023,
Seredynski:Kulpa-Hoffman-Deserable-2023,ACRI2024}.
In contrast,  the PD game is used here only to define a
the utility function 
that in turn defines the fitness and structure of a pattern.

\COMMENT{
\bibitem{Katsumata2008} 
Katsumata, Y., Ishida, Y., On a membrane formation in a spatio–temporally gener-
alized Prisoner’s Dilemma, Cellular Automata, Umeo, H., Morishita, S., Nishinari,
K., Komatsuzaki, T., Bandini, S. (eds) LNCS 5191 (2008) 60–66

\bibitem{Tretyakova2014} 
Tretyakova, A., Seredynski, F., Bouvry, P., Cellular automata approach to max-
imum lifetime coverage problem in wireless sensor networks, Cellular Automata,
W¸as, J., Sirakoulis, G., Bandini, S., (eds.) LNCS 8751 (2014) 437–446

\bibitem{Wolfram}
Wolfram, S. A New Kind of Science; Wolfram Media: Champaign, IL, USA, 2002; Volume 5.

\bibitem{Holland1992}
Holland, J.H. Genetic algorithms. Sci. Am. 1992, 267, 66–73.

\bibitem{Holland1995}
Mitchell, M. Genetic algorithms: An overview. In Complex; Wiley: Hoboken, NJ, USA, 1995; Volume 1, pp. 31–39

\bibitem{Holland2012}
Holland, J.H. Genetic algorithms. Scholarpedia 2012, 7, 1482.

\bibitem{Beasley1993}
Beasley, D.; Bull, D.R.; Martin, R.R. An overview of genetic algorithms: Part 1, fundamentals. Univ. Comput. 1993, 15, 56–69.

\bibitem{Hoffmann2023Loop}
Hoffmann, R.,
Generating Loop Patterns with a Genetic Algorithm and a Probabilistic Cellular Automata Rule.
Algorithms 16.7 (2023): 352.
}

\textit{Cellular Automata}. A CA is a grid of automata (cells). Every cell changes its state
depending on its own state and the state of its local neighbors according to a local rule.
The model of computation is parallel, simple, powerful, and has a wide range of applications 
\cite{WolframNewKind}.

\textit{Genetic Algorithm (GA) for Pattern Generation.}
We used already a GA to generate patterns with a global fitness function based on local conditions and local pattern matching
\cite{Hoffmann-2002-LoopPattern}.
GA is a generally accepted method for optimization. It dates back to John H. Holland 
\cite{Holland1992}, 
some overviews are presented in
\cite{
Beasley1993,
Holland2012,
Mitchell1995}.
We just used the classical techniques (crossover, mutation,
selection) in a simple way, i.e., we randomly selected the second parent (the mate) without
giving preference to parents with high fitness. This was sufficient for our purpose and it
was not the intention to find a more efficient evolutionary algorithm or heuristic.
This technique of finding optimal patterns with respect to a high rate of given local configurations  was also
used in previous works 
\cite{2014-Hoffmann-ACRI-Pattern,
2016-Hoffmann-Polonia-PathPattern,
2016-Hoffmann-D-ACRI-LinePattern,
2017-Hoffmann-D-PACT-MaxDomino-Agents}
in which finite state machines for moving agents were evolved
aiming at certain  pattern structures.

\COMMENT{
\bibitem{Beuchat2000}
Beuchat, J-L., and Jacques-Olivier Haenni. 
Von Neumann's 29-state cellular automaton: a hardware implementation.
 IEEE Transactions on Education 43.3 (2000): 300-308.

\bibitem{Mitchell1996}
Mitchell, Melanie, James P. Crutchfield, and Rajarshi Das. 
Evolving cellular automata with genetic algorithms: A review of recent work.
 Proceedings of the First international conference on evolutionary computation and its applications (EvCA’96). Vol. 8. 1996.

\bibitem{2014-Hoffmann-ACRI-Pattern}
Hoffmann, R. 
(2014).
How Agents Can Form a Specific Pattern.
ACRI Conf. 2014, LNCS 8751, pp. 660-669

\bibitem{2016-Hoffmann-Polonia-PathPattern}
Hoffmann, R. 
(2016).
Cellular Automata Agents form Path Patterns Effectively.
Acta Physica Polonica B Proceedings Supplement, Vol. 9 (2016) No.1

\bibitem{2016-Hoffmann-D-ACRI-LinePattern}
Hoffmann, R. and  D{\'e}s{\'e}rable, D.
(2016).
Line Patterns Formed by Cellular Automata Agents.
ACRI Conf. 2016, LNCS 9863, pp. 424-434

\bibitem{Bandini2009}
Bandini, Stefania, et al. 
Cellular Automata Pattern Recognition and Rule Evolution Through a Neuro-Genetic Approach.
J. Cell. Autom. 4.3 (2009): 171-181.

\bibitem{Capcarrere2002}
Capcarrere, Mathieu. 
Cellular automata and other cellular systems: design & evolution. No. 2541. EPFL, 2002.
 }

\textit{Finding a CA rule.}
The difficulty is to find an effective and efficient CA rule that can solve a 
given global problem/objective,
like density classification or
natural phenomena.
This issue is also called ``global-to-local problem''
\cite{Nagpal2008,Yamins2008}.
There are different ways to find a rule, such as
\begin{itemize}
	\item 
	Designing it by hand, for instance constructing self-replication rules
\cite{Beuchat2000}, or by discretizing differential equations. 

  \item
	Finding it with an evolutionary algorithm, like it was done in
   \cite
  {Mitchell1996,
  Capcarrere2002,
  2014-Hoffmann-ACRI-Pattern,
  2016-Hoffmann-D-ACRI-LinePattern}.  
	
	\item
	Designing it in a methodical way, 
  
          \begin{itemize}
            \item 
             partially by hand and partially tool-supported.
            This work belongs to this category.
              \item
              Or mainly automated using a tool-chain, like in
              \cite{Nagpal2008,Yamins2008,Bandini2009}.   
          \end{itemize}
	
\end{itemize}

In our previous work 
\cite{
2019-pact-domino,
2020-HoffmannSeredynski-SensorPoint,
Hoffmann-2021,
Deserable2023,
Hoffmann-2021-MinimalDominoPact,
Hoffmann-2022-PointPattern,
Hoffmann-2002-LoopPattern}
CA rules were designed in a methodical way
using templates (local matching patterns).
The templates are derived from a given set of tiles that define the pattern class.
The tiles are allowed to overlap, therefore
the number of tiles can vary and  the evolved patterns 
can easily adapt to different boundary structures.
In \cite{
Hoffmann-2022-PointPattern}
so-called ``\textit{point patterns}'' are evolved under cyclic boundary conditions.
These patterns contain 
points 
($\substack{000
          \\010
          \\000}$)    
which cover the 2D space with
a density that is not fixed, i.e.
items of the zero-hull may overlap. 

In this work, on the contrary, we shall evolve \textit{dense} patterns 
of points (for even grid sizes), and points mixed with dominoes (for odd grid sizes).
The concept of overlapping tiles is not used here, the patterns are evolved
only on the basis of templates which are conditions that strictly have to be
fulfilled in the final patterns.

\section{The problem and its solving method}

\subsection{The problem}


We assume a 2D grid of cells/agents $(i,j)$ with $i,j = 0~...~n-1$ and $N=n^2$.
Each cell has a state (color) $a_{ij} \in \{0,1\}$.
The array of cell states defines a \textit{field/pattern/configuration} $A = (a_{ij})$.
We use the following definitions and notations:

\begin{itemize}
	\item  

	${a}^{+K}_{ij} = (a_{ij}^0, a_{ij}^1,\ldots, a_{ij}^k,\ldots, a_{ij}^{K-1})$   

  
  The \textit{neighbors' states} (or the \textit{neighborhood configuration}) of $(i,j)$.

  The number of neighbors is $K$.
  The $k^{th}$ neighbor's state is denoted by $a_{ij}^k$.
  In addition, the relative position of each neighbor $k$ has to be declared.
  Usually the own cell's state $a_{ij}$ is included, then we can
  declare $a_{ij}^0 = a_{ij}$.
  But it is possible to exclude it, then only the states of the \textit{outer}
  neighborhood 
  are considered.   
       
       \begin{itemize}
         \item \textit{Remark}.
          Another way to define the neighborhood configuration of $a_{ij}$ is
          
          $a_{ij+P}=a_{ij+(P_0,P_1, .. P_k .. , P_{K-1}})$
          
          where $P$ is the \textit{neighborhood index vector}, with the elements (pointers, vectors)
          ${P_k}$ defining the relative position $(\Delta i, \Delta j)$ of the $k^{th}$ neighbor. 
          Then $a_{ij+P_k}=a_{ij}^k$.
       \end{itemize}

	\item
      $u: \{0,1\}^{K} \rightarrow [u_{min}~ .. ~u_{max}]$ 
      
      
      $u_{ij}= u(a^{+K}_{ij})$   
      
      The \textit{utility function}, or \textit{local value function}.	 
      The range is an interval in the set of real or integer numbers. 

   \item
      $V(A)= \sum_{\forall ij} u_{ij}$   
      
      The (global) \textit{Value} $V(A)$ of a pattern $A$ is the sum of all the cells' utilities.

   \item  
    $W(A)=V(A)/N$   
  
    The \textit{Wealth} $W(A)$ of a pattern $A$, that is the global value normalized to the
    number of agents, i.e. the agents personal utility/income when the global value is totally socialized,
    i.e. the average utility of a cell.  

   \item   
      
    \textbf{\textit{Maximize  W(A)}}
    
    The \textit{Objective Function} is $W$ to be maximized.    

    The wealth $W(A)$ of the whole community of agents is the objective function of our problem. 
    Alternatively the value $V(A)$ can be used if we do not need to compare patterns of different sizes $N$. 

\end{itemize}

\subsection{The problem-solving method}

This is an attempt to solve the problem with a probabilistic Cellular Automata (CA) rule as used before in
\cite{
2019-pact-domino,
2020-HoffmannSeredynski-SensorPoint,
Hoffmann-2021,
Hoffmann-2021-MinimalDominoPact,
Hoffmann-2022-PointPattern,
Hoffmann-2002-LoopPattern}.
Such a rule uses a list of so-called \textit{templates} $B_i$ that are local matching patterns,
often of size $3\times 3$ or $5 \times 5$.
If a template matches at $(i,j)$ in the aimed pattern $A$, then $a_{ij}$ is kept.
If a template matches except its center, then $a_{ij}$ is set to the center value of the template.
Otherwise $a_{ij}$ is changed randomly. 

In order to solve a given problem by this method, we need appropriate templates.
Now the idea is to look first at optimal solutions and then extract templates from them by the use of a gliding window.
How to find optimal solutions? 
A tool can be applied which was  used already in a former works.
This tool finds optimal patterns by a Genetic Algorithm for any global fitness that has to be defined.
After having found optimal patterns for small field sizes, templates can be extracted.
These templates are then inserted into a general CA rule with the expectation that optimal patterns can be evolved, also for larger sizes and different boundaries.

The problem depends heavily on the utility function being used, which has to be defined first.
The used approach allows to use any utility function, but the success of it can not be guaranteed.
Further research is needed to show the possibilities and limits. 
In our test case, this approach was successful.

In summary, our approach  consists of the following steps.

\begin{enumerate}
	\item
   \textbf{(Utility Function)} Define a utility function for the local neighborhood of cell states.

  \item
  \textbf{(Optimal Solutions)} Find optimal patterns for a small pattern size, using a GA or other optimization techniques.
  
  \item 
  \textbf{(Find Templates)} Analyze the optimal patterns found in step 2 using a gliding window. Each local configuration within the window is a candidate for a template.
  In the list of observed templates duplicates are deleted and missing symmetric ones are added.
  Remark: We assume that all symmetric templates are allowed in our problem, 
  i.e. the class of templates to which a template belongs shall be complete under
  all symmetries.  
  
  \item 
  \textbf{(Build CA Rule)} Insert the templates into a general CA rule that can evolve patterns by
  increasing the number of matching templates. 
  
  \item 
  \textbf{(Evaluation)} Compute the value of the evolved patterns. Evaluate if the values are optimal
  for patterns of sizes equal or different to the small size used in step 2. 
  In order to test if the evolved patterns are optimal,
  the optimal fitness of larger patterns have to be known 
  (Here we could find formula for that issue.)
  In general it is not easy to determine the value of a large pattern.

\end{enumerate}  

\section{A special utility function: the payoff of a game}

Our test case uses a utility function inspired by and derived from the Prisoner's Dilemma (PD) game.
The PD game is a version in the 2D space with several players in the neighborhood, the so-called SPD (\textit{spatial} PD) game.
Note that we do not use the PD game for any gaming or as a method for optimization,
like e.g. in
\cite{Katsumata2008,
Wang2022OptimizationGame}.
Here we just want to use the payoff matrix of the game as as an example
for the definition of the global pattern's value $V(A)$.
We define  $V(A)$ as the the sum of the local payoffs (utilities) that
all the agents receive together.  

In some precedent experiments it turned out that 
the usual PD parameters lead to the trivial pattern all 0 (all are agents are cooperating).
But when the parameters were changed, optimal patterns became a certain mixture of 0s and 1s 
(cooperators and defectors in terms of the game) that we aim at.

Now we will shortly explain the game and how the utility function is computed.

\subsection{Prisoner's dilemma game with two players}

The Prisoner's Dilemma is a well-known and well-studied game. 
In its basic form there are two players or agents, $X$ and $Y$.
They play with each other, and for each player the other player is its
opponent.
Both players choose simultaneously an individual state (also called action), 
that is either \textit{C(ooperate)} or \textit{D(efect)}.
After a play we have the combined state/action $(a_X,a_Y)$.
Now this tuple is evaluated using a payoff table which
defines the payoffs $(g_X,g_Y)$ for both players and for all combinations
of actions.

~~~$(g_{X}, g_{Y}) =
\begin{cases}
  (R,R) & \text{if~ } (a_{X}, a_{Y}) = (C,C) \\                      
  (S,T) & \text{if~ } (a_{X}, a_{Y}) = (C,D) \\                      
  (T,S) & \text{if~ } (a_{X}, a_{Y}) = (D,C) \\                      
  (P,P) & \text{if~ } (a_{X}, a_{Y}) = (D,D) \\
\end{cases}$

If we are only interested in the payoff of the player \textit{X},
depending on its own state $a_{X}$ and on the opponents' state $a_{Y}$
then we get, formulated in a different but semantically equivalent way:

~~~$g_{X}(a_{X}, a_{Y})  =
\begin{cases}
  R & \text{if~ } (a_{X}, a_{Y}) = (C,C) \\                      
  S & \text{if~ } (a_{X}, a_{Y}) = (C,D) \\                      
  T & \text{if~ } (a_{X}, a_{Y}) = (D,C) \\                      
  P & \text{if~ } (a_{X}, a_{Y}) = (D,D) \\
\end{cases}$

~~~$g_{X}(a_{X}, a_{Y})  =
\begin{cases}
  R/S & \text{if~ } (a_{X}, a_{Y}) = (C,C/D) \\                                          
  T/P & \text{if~ } (a_{X}, a_{Y}) = (D,C/D) \\                      
\end{cases}$

$
~~~g_{X}(a_{X}, a_{Y})  =
\begin{cases}
  g_{X}(a_{X}=C) = \begin{cases}
                            R & \text{if~ } a_{Y}=C\\
                            S & \text{if~ } a_{Y}=D\\
                           \end{cases}\\
                              
  g_{X}(a_{X}=D) = \begin{cases}
                            T & \text{if~ } a_{Y}=C\\
                            P & \text{if~ } a_{Y}=D\\
                           \end{cases}                      
\end{cases}
$

The payoff  depends on the parameters 
\textit{T, R, P, S,}
staying for \textit{Treason/Temptation, Reward, Punishment, Sucker}.
For the classical PD the relation

~~~$T > R > P > S$ and  $2R > (S+T)$

is required. The condition $2R >T$ ensures that playing $(C, C)$ gives
the maximum of $2R$ if the agent's and the opponent's payoff $(w = g_X + g_Y)$ are added.
Divided by $N=2$ (number of agents) gives the average (socialized/shared) ``Wealth'' \textit{W}$=w/N$ of each member of this small group of $N=2$ agents.

\textit{Remark}:
If the players have no information about the opponents playing strategy
then it would be most beneficial for both players to play $(D, D)$ which is called
a Nash Equilibrium (NE). Here we do not aim at a NE.

\subsection{Spatial PD game with K opponents}

We assume a 2D grid where at each point $(i,j)$ there is a player
who plays the PD game with his $K$ neighbors/opponents. 
Then a player receives a payoff $g^{k}_{ij}$ for the game with the $k^{th}$ neighbor, the total payoff is 

~~~$u^{total}_{ij} = \sum_{k=0}^{K-1} g^{k}_{ij}(a_{ij},a_{ij}^k)$~.   

where the payoff from the game with  opponent $k$ is

~~~$g^{k}_{ij}(a_{ij},a_{ij}^k)=
\begin{cases}
  R/S & \text{if~ } (a_{ij}, a_{ij}^k) = (C,C/D) \\                                          
  T/P & \text{if~ } (a_{ij}, a_{ij}^k) = (D,C/D)~. \\                      
\end{cases}$

We divide the total payoff  by the number $K$of opponents in order to normalize it, as we assume that 
the total wager is one unit only which has to be divided by the number of games/opponents.
The result is the agent's utility $u_{ij}$.

~~~$u_{ij} =u^{total}_{ij} /K $~.

We can distinguish two cases: \textit{with} and \textit{without self-play}. 
In the case \textit{without self-play} an agent plays only with others from the outer neighborhood.
In the case of \textit{self-play} an agent plays also with itself.

In the following we want to use a simplified game 
with the parameters (except \textit{T}) set to $(R=1), (P=S=0)$.
According to 
\cite{Nowak1993} 
the features of the game are maintained.

Now we compute the utility for the SPD game.
The payoff for a basic game  becomes $g(a_{Player})=R$  if $\COMMENT{(a_{Player},a_{Opponent})=} (C,C)$,
and $g(a_{Player})=T$ if $\COMMENT{(a_{Player},a_{Opponent})=} (D,C)$,
because the payoff is zero for $(C,D)$ and $(D,D)$.

$u^{total}_{ij} = ( R[a_{ij}=C] + T[a_{ij}=D] )N_C$, where $N_C= \sum_{k=0}^{K-1} [a^{k}_{ij}=C]$

We have used the notation $[condition]$ which evaluates to 1 if the condition is true, else to 0.
We declare $C=0$ and $D=1$, then we get $a_{ij}=[a_{ij}=D]$ and $\overline{a_{ij}}=(1-a_{ij})=[a_{ij}=C]$, 
and with $(R=1)$ the formula becomes

~~~$u^{total}_{ij} = (\overline{a_{ij}}+ Ta_{ij})N_C$~.

How to choose the parameters?
The condition for the classical PD is now 

~~~$(2R > (S+T)) = (2 > T > 1)$ for $R=1,S=0$.

This condition guarantees that the Wealth of the community is equal to 1 if all agents cooperate. 
Here we violate this condition by intention and change it to  $T>2$, because we aim at patterns with a Wealth $W>1$.
Such patterns will be a structured mixture of cooperators and defectors.
We exclude $T=2$ because then the chosen actions would not matter and 
the Wealth would be 1 independently of the pattern. 

  In our application we use the spatial PD game with  $K=9$ neighbors
  out of the MOORE neighborhood (with self-play), and the parameters are set to
  $T=3, R=1, P=0, S=0$.
  A defecting agent with 8 cooperating neighbors yields maximal a payoff of 24.
  A cooperating agent with 9 cooperating neighbors yields maximal a payoff of 9.
  The boundaries are  
  
  ~~~$0 \leq u^{total}_{ij} \leq T \times (K-1) = 3 \times 8 = 24$ 
  
  ~~~$0 \leq u_{ij} \leq  T \times (K-1)/K = 24/9=2.67$~.

\subsection{The pattern's value for the utility of the PD game}


The \textit{Total Payoff Sum}  that all $N$ agents receive together is

~~~\textit{TPS} = $\sum_{\forall ij}{u}_{ij}^{total}$~, where ${u}_{ij}^{total}$ is the player's total payoff. 

Dividing by the number of opponents gives the \textit{Value} of the pattern $A$

~~~$V(A) = TPS/K = \sum_{\forall ij}{u}_{ij}$~.

Further dividing by the number of agents gives the \textit{Wealth}
which is the income of each agent when all payoffs are normalized 
and shared (socialized):

~~~$W(A) = V(A)/N$~.

\subsection{Estimation of the maximal reachable wealth}

We want to estimate the wealth depending on (1) the rate of cooperators rsp. defectors in the
whole population, and (2) the parameters $T,R,P,S.$
Agents are in the cooperating state $C$ with a rate of $\pi_C$,
and in the defecting state $D$ with a rate of $\pi_D=1-\pi_C$.

In order to compute the wealth, we can assume that
$N \pi_C$ cooperators and $N \pi_D$ defectors
play with all $N$ agents as opponents. 
If the total payoff of an agent is divided by the number of $N$ of opponents,
then we get the normalized total payoff 
$\textit{payoff}(a_{ij})=u_{ij}=u_{ij}^{total}/K$. 
We declare

~~~$\textit{payoff}(D) = u_{ij}(a_{ij}=D)$ ,  $\textit{payoff}(C) = u_{ij}(a_{ij}=C)$ .

The \textit{expected wealth W} is the sum of
(i) the  $\textit{payoff}(D)$ which a player in state $D$ receives, multiplied by the rate of occurrence $\pi_D$, and 
(ii) the  $\textit{payoff}(C)$ which a player in state $C$ receives, multiplied by the rate of occurrence $\pi_C$.

~~~$W= \pi_D \times \textit{payoff}(D) + \pi_C \times \textit{payoff}(C)$ , with

~~~$\textit{payoff}(D)= P \pi_D + T \pi_C$ , and

~~~$\textit{payoff}(C)= R \pi_C + S \pi_D$ .

\begin{figure}[H] 
\centering
(a)\includegraphics[width=5cm]{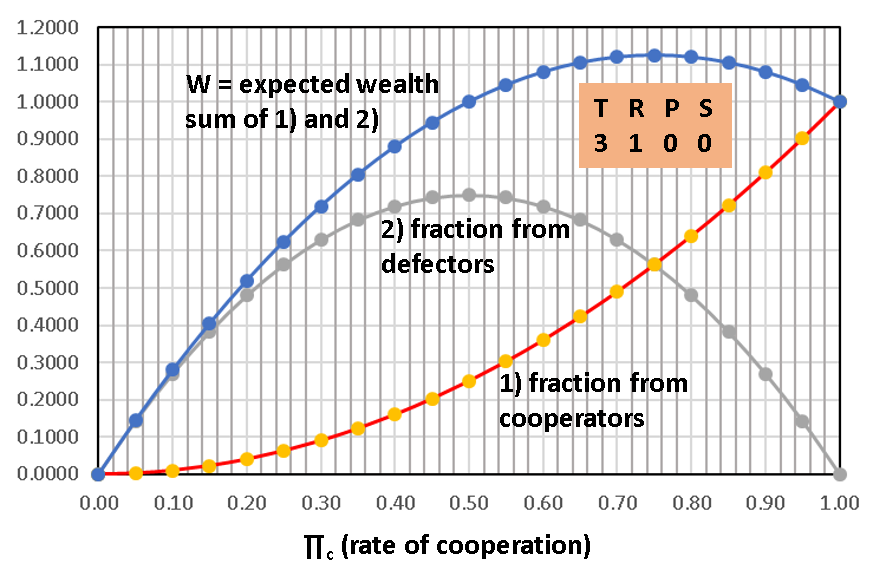}
(b)\includegraphics[width=5cm]{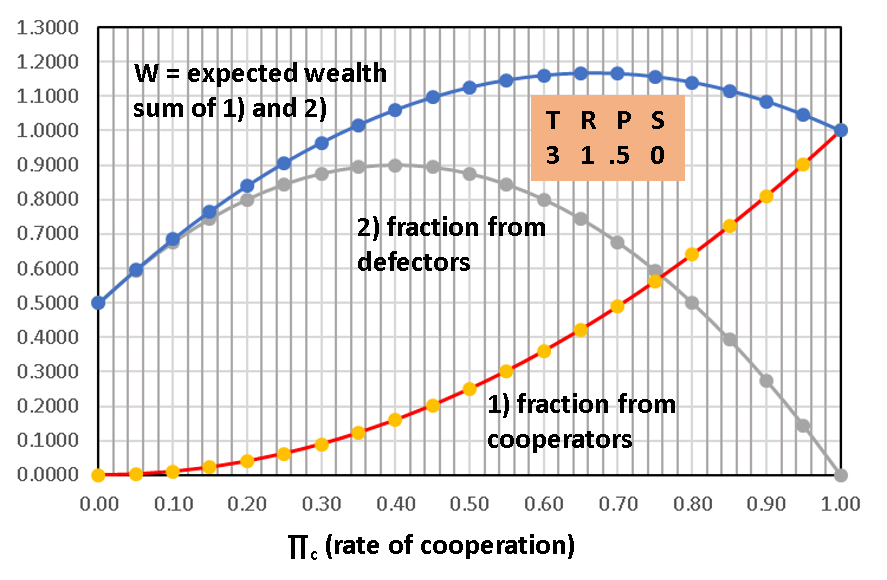}		

(c)\includegraphics[width=5cm]{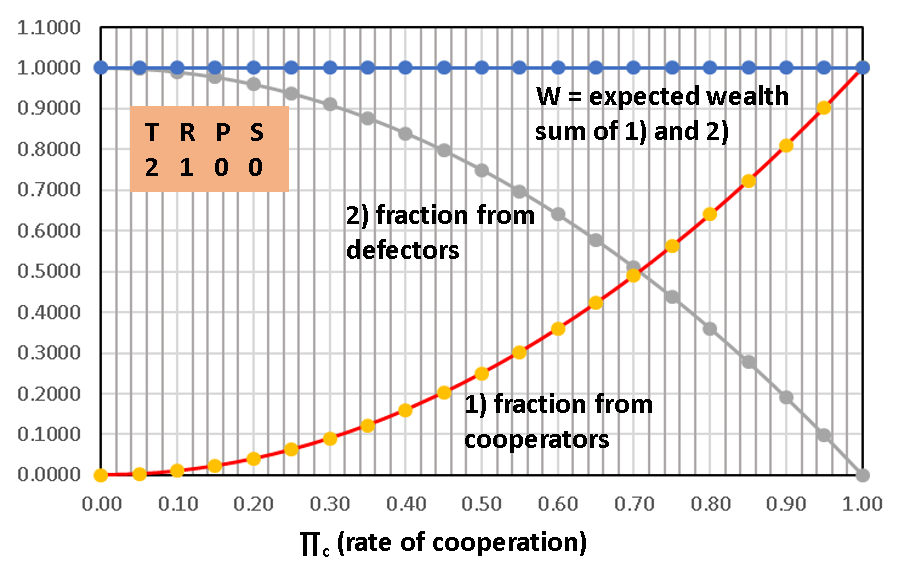}
(d)\includegraphics[width=5cm]{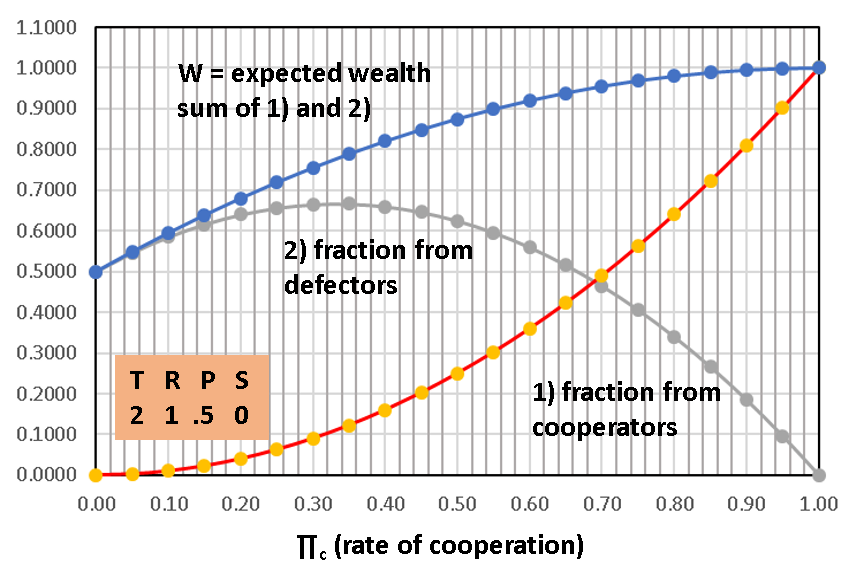}

(e)\includegraphics[width=5cm]{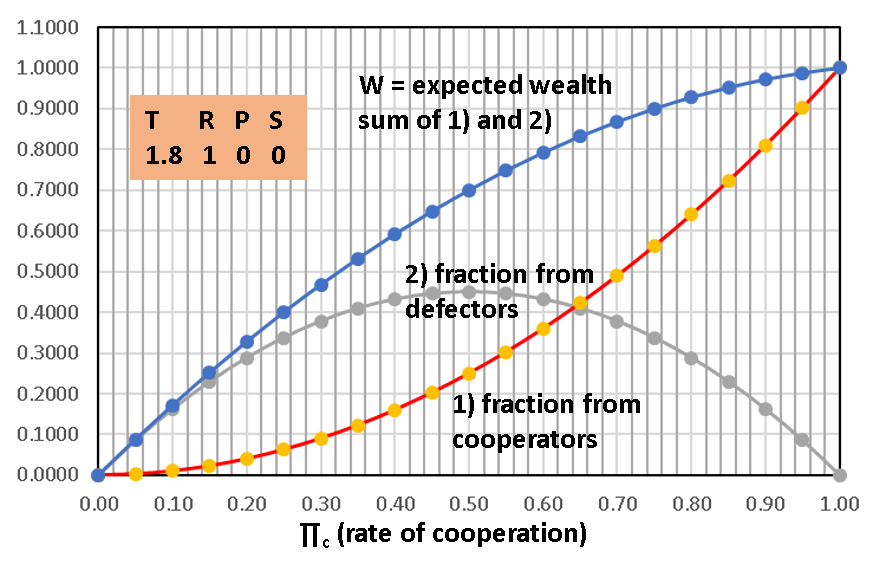}
(f)\includegraphics[width=5cm]{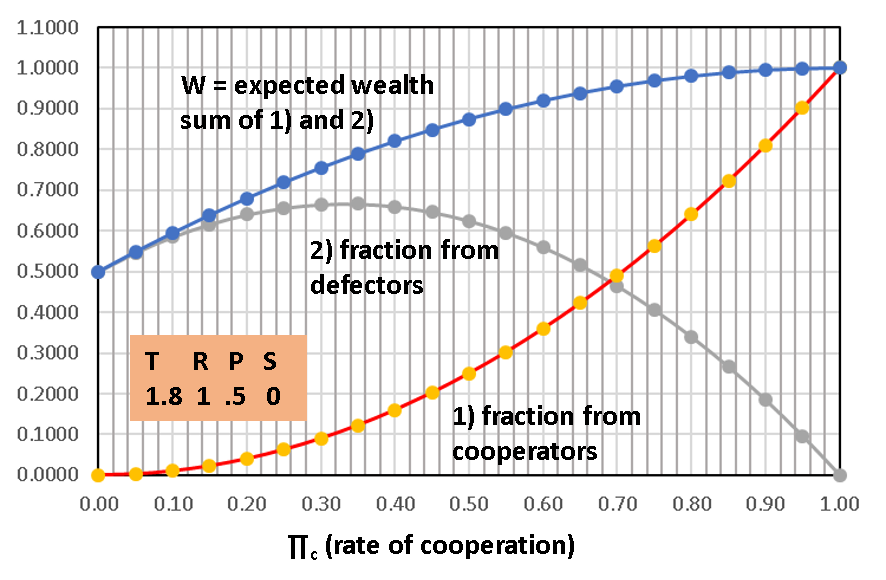} 	
\caption{
The expected wealth vs the rate of cooperation, for different
parameter settings $(T,R,P,S)$.
The parameters shown for case (a) are used in our problem.
}
\label{payoff_graph1}
\end{figure}

Fig.~\ref{payoff_graph1} shows the expected wealth for different parameter settings.
(a) This case is the one treated in this work. 
It shows a concave maximum with 1.125 at  $\pi_C=0.75$.
This approximation relates very good to the real optimal value (1.1944) obtained by the optimization procedures
in the following sections. 
Note that the influence of the used $3 \times 3$ MOORE neighborhood was not taken into account.
(b) This graph starts at $P=0.5$ and has its maximum with 1.166 at 0.65.
(c) The wealth is constantly one, this parameter setting is not a case for optimization.
(d -- f) These settings are classical and are shown for information only. 
The maximal wealth increases steadily with the rate of cooperation 
and reaches 1 at $\pi_C=1$.

\section{Using a Genetic Algorithm to find optimal patterns}

\begin{figure}[H]

\begingroup
\parindent0mm  
\hrulefill

\textbf{Algorithm 1: Generating a Binary Pattern with a Genetic Algorithm}

\vspace{-6pt}
\hrulefill

\par  

\begin{tabular}{lll} 
Types:          & $Pattern = array [0 ~..~ n_1-1, 0~..~ n_2-1] ~\textit{of}~ \{0,1\}$  \\
                & $Solution = (Pattern, Fitness)$ \\
Data \& Output: & \textit{S is an array [1 ~..~ M] of Solution}  \\
Temporary:      & \textit{Offspring is a Solution}\\
\end{tabular}

\noindent\hrulefill 

%
$(1)~ S.Pattern \leftarrow RandomPatterns$ 

\begin{itemize}
\item []  (2) \textbf{while not} $TerminationCondition$ 
   
    \begin{itemize}  
    \item[] (3) \textbf{foreach} $S_i$ \textbf{in} $S$      
    
    ~~~~~~(4) \textit{Select randomly a Mate} $S_j$     

    ~~~~~~(5) \textit{Offspring.Pattern} = \textit{Mutate(Crossover}($S_i,S_j)$)
      
    ~~~~~~(6) \textit{Offspring.Fitness} $\leftarrow$ \textit{fitness(Offspring.Pattern)} 
  
    ~~~~~~(7) \textbf{if~}\textit{Offspring.Pattern}  $\notin$ \textit{S.Pattern } 
    
    ~~~~~~~~~~~~~~~~~~~~~~\textbf{and}  \textit{Offspring.Fitness $> S_i$.Fitness}
    
    ~~~~~~~~~~~~\textbf{then} $S_i \leftarrow$ \textit{Offspring}

    \item [] (3) \textbf{end foreach}      
    \end{itemize}
   
\item []  (2) \textbf{end while}
\end{itemize}

(8) $S \leftarrow SortByFitness(S)$

\hrulefill
\endgroup

\caption{Algorithm that generates optimal patterns with respect to a global 
fitness function. 
The fitness function used here is the \textit{Wealth} of the community, taking into account 
the local payoffs of all agents. 
}

\label{Algorithm1}
\end{figure}

In a first step we use a GA to find optimal patterns with
respect to the defined global fitness function.
Such patterns will be used 
(a) as master solutions from which local matching templates are extracted
to be inserted in a CA rule,
and 
(b) for comparison with the fitness of the later CA evolved patterns. 

The \textit{fitness} function $f(A)$ is computed as follows. 
At every site of the pattern the total payoff from the SPD game is computed and then
summed up over all cells. It is

~~~$f(A) =  \sum_{\forall{ij}} {\textit{u}_{ij}^{total}} =\textit{TPS} = W(A) \times N \times K$~.

The algorithm used is given in Fig.~\ref{Algorithm1}.
A possible solution (an individual) is a tuple $(Pattern, Fitness)$.
An array/list of $M$ solutions  is the data and output that is manipulated  by the algorithm.
(1) The pattern list is initialized randomly. 
(2) The while-loop is repeated until a termination condition becomes true, i.e. 
the number of iteration has reached a certain limit, and/or
another global condition, like a certain fitness level or certain pattern features.
(3) 
In the for each-block better solutions are searched for.
For each current existing individual (tentative solution)
an offspring (a new individual) is generated by crossover and mutation.
The current individual is replaced by the offspring if it has a higher fitness. 
(4) A mate  $S_j$ is selected at random from the list. 
(5) A new offspring pattern is computed by crossing over $S_i$ with $S_j$,
and then applying a mutation. 
(6) The fitness of the offspring pattern is computed and stored within the offspring.
(7) The offspring replaces the current individual $S_i$
if its fitness is higher.
(8) The list of solutions is sorted for output.

The used GA is a simple form using the classical GA algorithm basics.
The goal was to generate optimal patterns in a simple way and not to 
optimize the algorithm itself which is a topic for further research. 
Only one list of individuals is used and not two (old and new generation).

Each individual is treated separately and is expected to improve by crossover
with any other individual, not depending on their fitness. 
Thereby a high diversity is supported though the speed of improvements may not be so high.  
Crossover is performed in the following way (uniform crossover with a certain probabilty):
(1) Each bit of the offspring $\textit{Crossover}(S_i, S_j)$
(without mutation) is taken from
the mate $S_j$ with probability $p_1$
otherwise unchanged from the parent $S_i$.
(2) Then mutation is performed on each bit with probability $p_2$ yielding 
$Mutate(Crossover(S_i, S_j))$.
Then the fitness of the offspring is computed and used for replacement
in the case of improvement. 

The used size of the population (the number of possible solutions) was 40,
and the probabilities  were set to $p_1= 0.2, p_2=0.05$.
Optimal or near-optimal solutions were found within a number of iterations between 1 000 and 10 000
in a few minutes 
on a desktop PC with an Intel Core i5-3470 CPU @ 3.20GHz and 8 GB RAM
under Windows 10. The programming language was Free Pascal using one thread only.

\section{Found optimal patterns}

We want to characterize a pattern \textit{A} by the characteristic 

~~~$CC=(W(A), f(A)=\textit{TPS(A)}; n, n^2, b, b/n)$, where

~~~~~~$b$ is the number of cells in state 1 (Defect).

\subsection{Patterns of size 6 x 6}

\begin{figure}[H] 
\centering

\includegraphics[width=11cm]{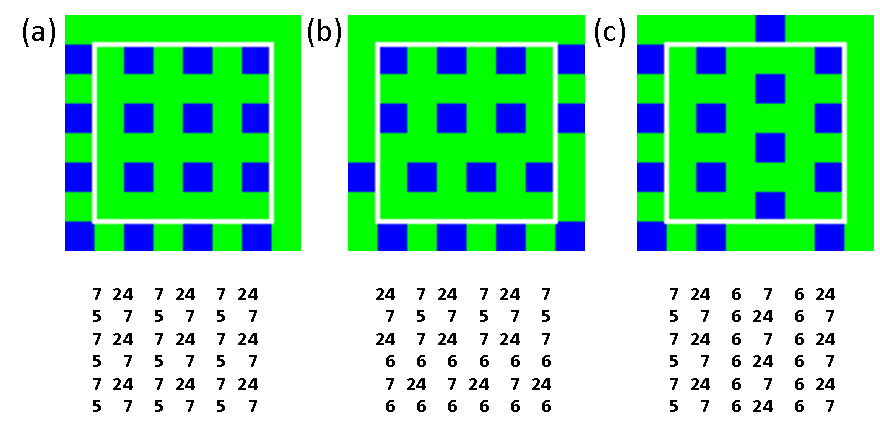}
\caption{
6 x 6 Optimal pattern evolved by a Genetic Algorithm.
The total payoffs are shown below the patterns. 
Fitness is 387, the maximum.
}
\label{ga6x6}
\end{figure}

The found optimal patterns for $n =6 \times 6$ are shown in Fig.~\ref{ga6x6}.
Their characteristic is 
$CC=(1.19444, 387; 6, 36, 9, 0.250)$. 
The payoffs of the cells are given below the patterns.
Summing up all payoffs gives the fitness of 387, which corresponds
to a wealth of 1.19444.
The average wealth of each agent is 19.444\% higher than the pattern all $C$ (all 0) which 
is the baseline wealth of 1 for comparison.
The sum of all payoffs is 387, and the number of \textit{Ds} (ones) in the pattern is 9 out of 36 (25\%).
The patterns contains a maximal number of \textit{points}, 
where a point is a '1' surrounded by 8 zeroes. 
We consider patterns \textit{equivalent} if they are equal under the symmetries shift, rotation,
and reflection.
The patterns Fig.~\ref{ga6x6}(b) and (c) are equivalents.

How equal is the community in terms of personal income/payoff.
If all agents would cooperate then the pattern is all 0, 
and every agent has a base income (total payoff) of 9.
(It becomes 1 if it is normalized through division by $K$).
In an optimal pattern there are winners (income higher than 9) and losers (income lower than 9).
For this example the winners are the defectors scoring 24, and the losers are the
cooperators scoring 5, 6 or 7. But as we assume that all profits are shared
the average income  becomes $TPS/n^2= W\times K = 10.75$, 1.75 units better than 
the base income. 

\begin{figure}[H] 
\centering
(1)

\includegraphics[width=11cm]{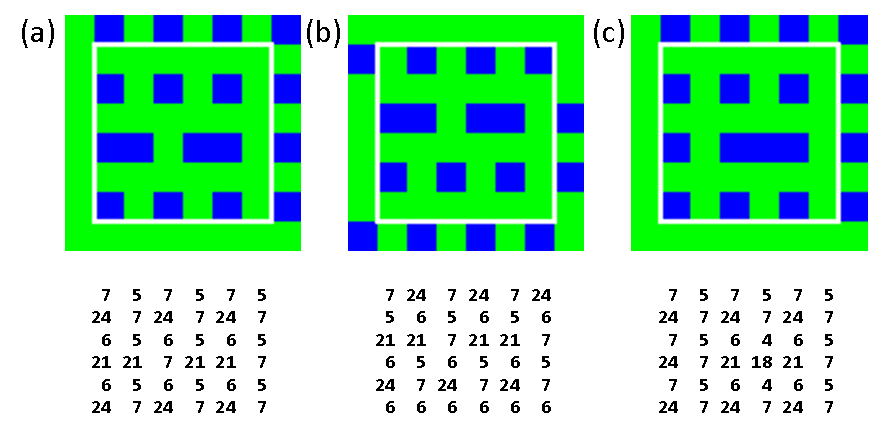}

(2)

\includegraphics[width=11cm]{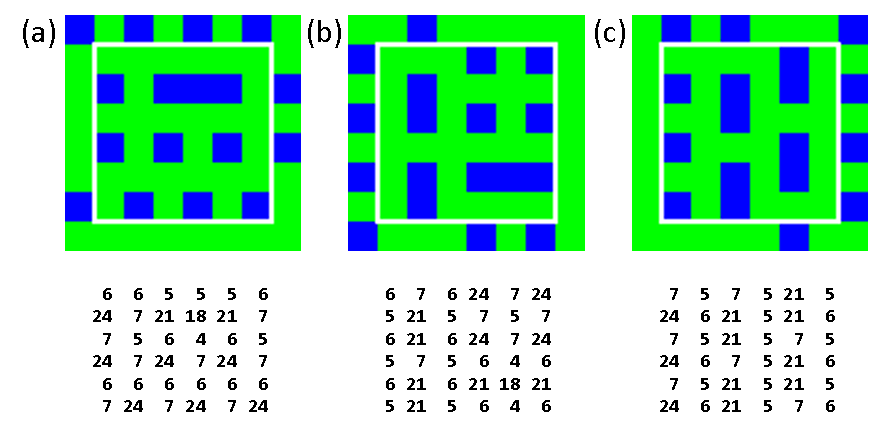}
 		 	
\caption{
6 x 6 Near-optimal patterns evolved by a GA.
Fitness is 386 for (1a)--(2a), and 385 for (2b)--(2c).
}
\label{ga6x6nonopt}
\end{figure}

Near-optimal patterns can be of interest, too.
The GA evolved  second best patterns as depicted in Fig.~\ref{ga6x6nonopt}.
Their characteristic is

~~~$CC=(1.19136, 386; 6, 36, 10, 0.278)$~.

It is remarkable that their fitness 
is only one resp. two  less then the maximum.
The number of points has incremented by one (from 9 to 10) compared to the optimum. 
Now the pattern contains not only  points but also dominoes (11-lines/rectangles) or 111-lines.

\subsection{Patterns of size 5 x 5}

\begin{figure}[H] 
\centering

\includegraphics[width=8cm]{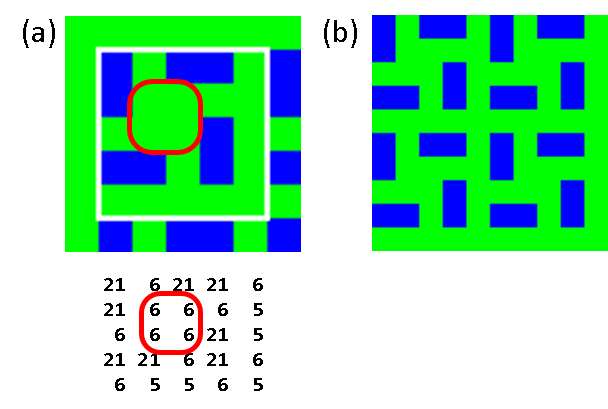}
\caption{
(a) 5 x 5 optimal pattern evolved by a Genetic Algorithm.
The total payoffs are shown below it. 
Fitness is 265 and wealth is 1.17778, the maximum.
Inside the red marked area there is a block of 4 zeroes.
(b)
The pattern (a) (inside the white marked boundaries)
is 4 times replicated, twice horizontal and twice vertical. 
This ``quad'' representation emphasizes the inherent structures.
}
\label{ppt-ga5x5opt}
\end{figure}

There is only one optimal pattern (Fig.~\ref{ppt-ga5x5opt}), suggested from many runs of the GA.
The characteristic is $CC=(1.17778, 265; 5, 25, 8, 8/25=0.32)$. 
The income of the defectors/winners with cell state 1 is 21, while that of the cooperators/losers is only 5 or 6.
Recall that each agent would have an income of 9 (the baseline), 
if all of them would cooperate (corresponding to an all zero pattern).
In this pattern there are 6 winners and 19 losers.
As we assume that the winners are social and altruistic they share their income with the whole community
aiming at an equal or fair (close to equal) distribution.
As the total income of the whole community is $TPS=265$, we divide it by the number of agents.
The result is a shared common income of $265/25=10.6$ which is an improvement of $1.6$ compared to the baseline of 9.
In order to motivate agents to become a dealer one could choose a distribution with partial sharing, for example
income 10 for a consumer, and income 12.5 for a dealer.

We observe in the pattern 4 dominoes arranged in a  ``cycle''.
And we find a square block of 4 consumers (marked in red) within this cycle. 
We call such a block of $2\times 2$ zeroes, framed/confined by certain neighbors
(typically  dominoes or points) 
in a $4\times 4$ window, a \textit{singularity}.
In other non-optimal patterns, and in optimal patterns of a higher odd size $n >5$, we can also find
singularities.

\begin{figure}[H] 
\centering

\includegraphics[width=\textwidth]{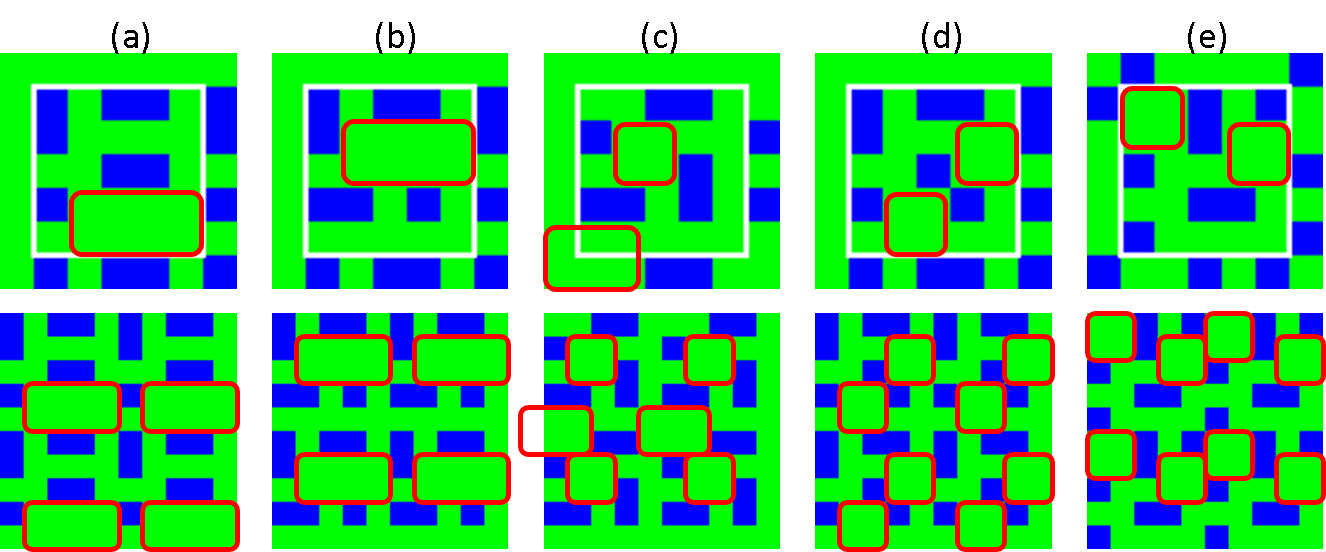}
\caption{
 $5 \times 5$ near-optimal patterns in normal and in quad representation.
Fitness is 262 (3 less than the maximum) and wealth is  1.16444.
The marked areas a small blocks of zeroes (cooperators/consumers).
}
\label{ppt-ga5x5optnear}
\end{figure}

Fig.~\ref{ppt-ga5x5optnear} shows some near-optimal patterns with a fitness of 262, three less then the optimum.
The characteristic is $CC=(1.16444, 262; 5, 25, 7, 7/25=0.28)$.  
We can observe patterns with (3 dominoes and 1 point), 
and (2 dominoes and 3 points where two of them are diagonally connected).
In these patterns we find blocks (marked by red rectangles) 
of consumers of size $2 \times 2$ or $2 \times 3$.
If we analyze the optimal pattern (Fig.~\ref{ppt-ga5x5opt}) with respect to such blocks,
we find just \textit{one} $2 \times 2$ zero-valued block in the middle of a ``circle'' of 4 dominoes.

\subsection{Patterns of size 7 x 7}

\begin{figure}[H] 
\centering

\includegraphics[width=8cm]{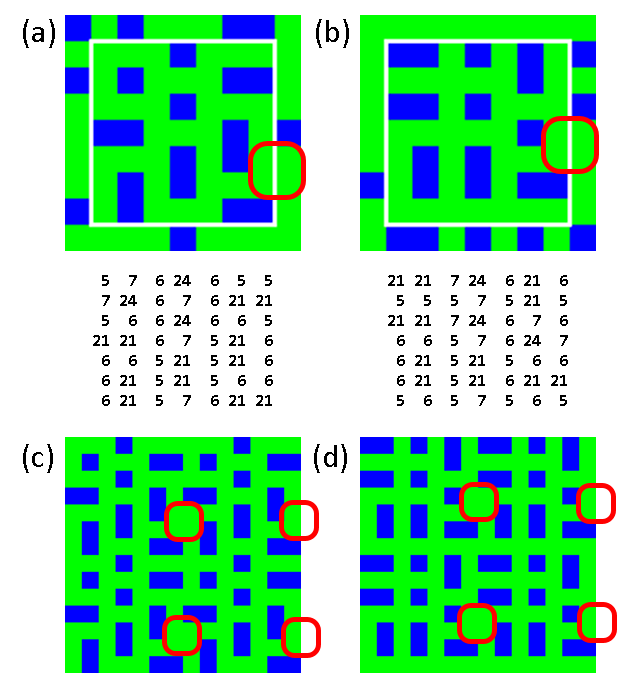}
\caption{
7 x 7 Optimal pattern evolved by a Genetic Algorithm.
The total payoffs are shown below the pattern (a). 
Fitness is 522 and Wealth is 1.18367, the maximum.
Patterns (c, d) are the quad representations of (a, b).
The marked areas are singular points of 4 zeroes. 
}
\label{ppt-ga7x7opt}
\end{figure}

The optimal $7 \times 7$ pattern (Fig.~\ref{ppt-ga7x7opt})
has the characteristic

~~~$CC=(1.18367, 522; 7, 49, 15, 15/49=0.306)$.

The pattern contains 6 dominoes and 3 points. 
We find exactly one singularity (marked). 

\subsection{Patterns' characteristics for odd $n>7$}

More optimal patterns were evolved by the GA for $n=9,11,13,15$.
Table \ref{TableDataOdd} gives their characteristic data.
All the given values increase with $n$, except the density $b/n^2$ of the ones which decreases. 

We can conclude  from the table (without giving a proof) that the following relations hold, for
$odd(n)$ and $n\geq 5$:

~~~$N_{domino}(n)=n-1$

~~~$N_{point}(n+2) = N_{point}(n)  + n-2$, with $N_{point}(5)=0$ 

~~~$N_{point}(n)= m+m(m+1)$, where $m= (n-5)/2$

~~~$b=2 N_{domino}(n) + N_{point}(n)$

~~~$density = b/n^2$

As shown later in Sect. 
\ref{Odd pattern sizes}
 we can also compute the total play sum \textit{TPS}
and the wealth \textit{W}:

~~~$\textit{TPS} = 265 + 128m + 43(m(m + 1))$

~~~$W= \textit{TPS}/n^2/(K=9)$.

For very large $n$ the wealth converges to 1.1944 and the density to 1/4,
the same values as for patterns of even size. 
This can also be explained by the fact that very large patterns are almost totally filled with points.  

\begin{figure}[H] 
\centering

\includegraphics[width=\textwidth]{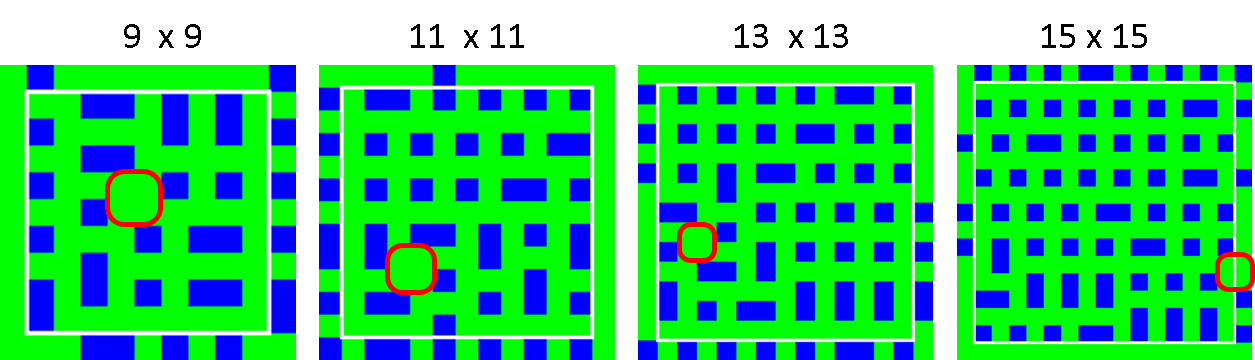}

\caption{
Optimal patterns for $n=9$ to $n=15$. 
}
\label{ppt-ga9bis15opt}
\end{figure}

\begin{table}[H] 
	\caption{
The patterns' characteristics for odd $n$.}	
\begin{center}
	\begin{tabular}{ccccccccc}
  

   &     &   & \textit{wealth} &\textit{TPS} & \textit{number} & \textit{number} & \textit{number}   & \textit{density}            \\
$m$ & $n$  & $n^2$ & $W$         &         & \textit{dominoes} & \textit{points}       & \textit{b of ones}  & $b/n^2$  \\				
				
\hline		   
-   & 3    &    9      &$     1.12346  $&  91   &  2      &   0      &    4  &$    0.444 $\\
0   & 5	   &	 	25	   &$     1.17778  $&  265  &  4      &   0      &    8  &$    0.320 $\\
1   & 7	   &    49     &$     1.18367  $&  522  &  6      &   3      &    15 &$    0.306 $\\
2   & 9    &    81     &$     1.18656  $&  865  &  8      &   8      &    24 &$    0.296 $\\
3   & 11   &    121    &$     1.18825  $&  1294 &  10     &   15     &    35 &$    0.289 $\\
4   & 13   &    169    &$     1.18935  $&  1809 &  12     &   24     &    48 &$    0.284 $\\ 
5   & 15   &    225    &$     1.19012  $&  2410 &  14     &   35     &    63 &$    0.280 $\\



  \end{tabular}
\end{center}
\label{TableDataOdd}
\end{table} 

\section{Extracting templates from the optimal patterns}

We want to use the found optimal patterns as masters and use their inherent local structures
to generate similar (hopefully optimal or near-optimal) patterns. 
In this second step we extract from the optimal patterns local matching patterns (templates) that we will
use later (in the third step) in a CA rule to generate patterns that are similar to the given optimal ones.

A special software tool was developed that can analyze a given pattern with respect to their inherent
local neighborhoods.
A small window (typically a Moore or von-Neumann neighborhood) is gliding over the whole pattern. 
All found neighborhood configurations (we call them \textit{found templates}) $T^{found}_i$ are stored in a list ($T^{found}_i$).
Multiples are deleted.
We check for symmetries (rotations, mirror against horizontal/vertical center line),
and then we form classes of symmetric templates. 
If a class is not complete, we add the missing symmetric templates.
We have assumed for our problem that each templates belongs to a certain class
(under symmetries) of templates, i.e. each template is a representative of a certain class
of equivalent templates. 

\textit{Remark}: In general, it is not necessary to complete the classes symmetrically, especially if we know
that some symmetric templates will never occur in the given pattern or a similar pattern that we
aim at and if we want to generate patterns on the basis of extracted templates only.

For our problem we assume that the inherent local structures can be expressed by templates of size $3 \times 3$.
Indeed, for our problem, this size seems to be sufficiently large, but this cannot be stated in general.
All extracted templates are shown in Tab.~\ref{TableTemplatesEven},
where T0--T7 define point-only patterns for an even field size $n$,
and T0--T51 the more complex patterns for $n$ odd.

\begin{table}[H] 
\caption{
T0--T7: These templates define the possible neighborhoods if the field size $n$ is even,
i.e. when then patterns contain points only. 
T0 represents a point (a one surrounded by zeroes).
T2 is a $90^\circ$ rotation of T1, T4--T7 is a class of equivalents under rotation.
T0--T51: The templates for an odd field size.}

\begin{center}
\footnotesize
\begin{verbatim}
            T0      T1      T2      T3      T4      T5      T6      T7 
            A       B0      B1      C       D0      D1      D2      D3
          0 0 0   0 0 0   0 1 0   1 0 1   1 0 1   0 1 0   0 0 1   1 0 0   
          0 1 0   1 0 1   0 0 0   0 0 0   0 0 0   0 0 0   1 0 0   0 0 1   
          0 0 0   0 0 0   0 1 0   1 0 1   0 1 0   1 0 1   0 0 1   1 0 0  
          
            T8      T9     T10     T11     T12     T13     T14     T15 
            E0      E1      E2      E3      F0      F1      F2      F3
          0 0 0   0 0 0   0 1 0   0 0 0   1 1 0   0 1 1   1 0 1   0 0 0
          1 1 0   0 1 1   0 1 0   0 1 0   0 0 0   0 0 0   1 0 1   1 0 1
          0 0 0   0 0 0   0 0 0   0 1 0   1 1 0   0 1 1   0 0 0   1 0 1
          
           T16     T17     T18     T19     T20     T21     T22     T23
            G0      G1      G2      G3      G4      G5      G6      G7
          1 1 0   0 1 0   0 1 1   0 1 0   0 0 1   0 0 0   1 0 0   0 0 0                                   
          0 0 0   0 0 0   0 0 0   0 0 0   1 0 1   1 0 1   1 0 1   1 0 1                                   
          0 1 0   1 1 0   0 1 0   0 1 1   0 0 0   0 0 1   0 0 0   1 0 0
          
           T24     T25     T26     T27     T28     T29     T30     T31                           
            H0      H1      H2      H3      H4      H5      H6      H7                                                           
          1 1 0   1 0 1   0 1 1   1 0 1   1 0 1   1 0 0   1 0 1   0 0 1                           
          0 0 0   0 0 0   0 0 0   0 0 0   0 0 1   0 0 1   1 0 0   1 0 0            
          1 0 1   1 1 0   1 0 1   0 1 1   1 0 0   1 0 1   0 0 1   1 0 1                                 
          
           T32     T33     T34     T35     T36     T37     T38     T39
            I0      I1      I2      I3      J0      J1      J2      J3 
          1 1 0   0 1 1   0 0 1   1 0 0   1 1 0   1 0 0   0 1 1   0 0 1                                 
          0 0 0   0 0 0   1 0 1   1 0 1   0 0 0   0 0 0   0 0 0   0 0 0                                 
          0 1 1   1 1 0   1 0 0   0 0 1   1 0 0   1 1 0   0 0 1   0 1 1                                                        
           
           T40     T41     T42     T43     T44     T45     T46     T47
            J4      J5      J6      J7      K0      K1      K2      K3                             
          1 0 1   0 0 0   1 0 1   0 0 0   0 1 0   0 0 1   0 1 0   1 0 0                                
          0 0 1   0 0 1   1 0 0   1 0 0   0 0 0   0 0 0   0 0 0   0 0 0                    
          0 0 0   1 0 1   0 0 0   1 0 1   0 0 1   0 1 0   1 0 0   0 1 0                                                         
                
           T48     T49     T50     T51
            K4      K5      K6      K7         
          0 0 0   1 0 0   0 0 0   0 0 1               
          0 0 1   0 0 1   1 0 0   1 0 0              
          1 0 0   0 0 0   0 0 1   0 0 0                                                          
\end{verbatim}
\end{center}
\label{TableTemplatesEven}
\end{table} 

\textbf{The templates for \textit{n} even.}
Let us extract the templates from the optimal patterns shown in Fig.~\ref{ga6x6}.
As the optimal patterns in Fig.~\ref{ga6x6} are simple structured we can find the templates also by visual inspection
without using a special tool.
We find easily the template T0  as a ``point'' surrounded by zeroes.
In addition we extract the templates T1, T2, T3 from Fig.~\ref{ga6x6}(a).
They define the neighborhood of a zero valued cell when two lines of points are not shifted against each other.
For point lines that are shifted against each other, we find T4, T5 in Fig.~\ref{ga6x6}(b), and  then T6, T7 in Fig.~\ref{ga6x6}(c).
So we have found all relevant templates for the optimal patterns shown in Fig.~\ref{ga6x6}. 
Note that other optimal patterns differ only slightly (lines of points can be shifted against each other),
but it is obvious that the set of found templates is complete in the sense that we can construct
any optimal pattern with them by overlaying them without conflict and thereby covering the whole area. 
Not all templates need to be part of an optimal pattern, for instance the pattern in
Fig.~\ref{ga6x6}(a) uses templates (T0--T3) only, 
and Fig.~\ref{ga6x6}(b) uses only the templates (T0--T3), T4, T5,
and Fig.~\ref{ga6x6}(c) uses only the templates (T0--T3), T6, T7.
We can also imagine an optimal pattern of size $8 \times 8$ which contains 4 rows that are shifted 
row by row. Such a pattern can be defined by the templates T0, T6, T7 only.

\textbf{The templates for \textit{n} odd.}
Optimal patterns of size $5 \times 5$ and   $7 \times 7$ were generated by the  GA pattern generator. 
Then several optimal patterns were analyzed by a tool using the gliding window technique.
All found templates were gathered and enhanced by missing symmetric ones.  
Then the list of templates was sorted into categories by hand as shown in Table \ref{TableTemplatesEven}.
Templates T0--T7 are the basic templates also used for patterns with even size. 
E0--E3 appear as result of dominoes fitting into  $3 \times 3$ target window (not all surrounding zeroes fit into that window). So we have it to do with partial neighborhoods which are induced by certain
local configurations of dominoes and points. 
F0--F3 arise as the result of two parallel dominoes. 
G0--G7 and H0--H7 arise as the result of one domino and one point (or alternatively a part of a domino).
I0--I3 arise as the result of two skewed dominoes.
J0--J7 and K0--K7 are derived from 
\textit{singularities} observed in optimal patterns.
A singularity is a $4 \times 4$ local pattern that includes a $2 \times 2$ square block of zeroes. 
We carefully inspected the optimal patterns generated by the GA and found exactly one singularity in each optimal pattern of odd size. 
A proof would be welcome giving evidence to this observation. 

\begin{figure}[H] 
\centering

\includegraphics[width=9cm]{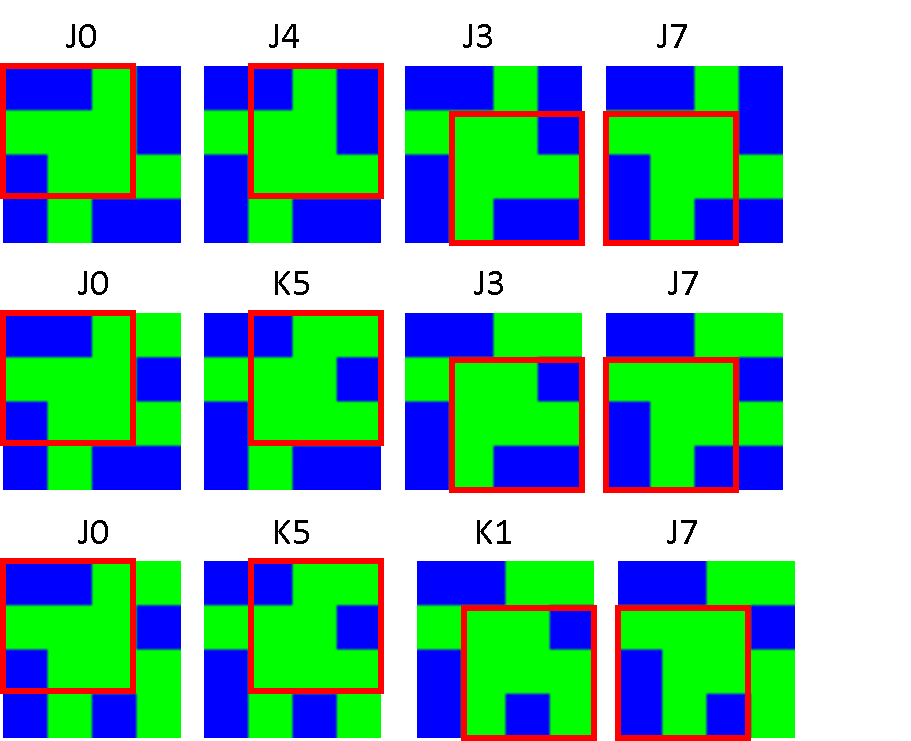}

\caption{
Three singularities of size $4 \times 4$ were detected in optimal $7 \times 7$ GA patterns,
shown in the three rows. 
A singularity is a $2 \times 2$ block of zeroes enclosed by its direct neighbors.
Templates of size $3 \times 3$ are derived from the singularities using a gliding window (marked in red).
We find the templates J0, J3, J4, J7, K1, and K5. The missing symmetric ones have to be added
resulting in J0--J7 and K0--K7.
}
\label{ppt-singularities}
\end{figure}

Fig.~\ref{ppt-singularities} shows the three detected singularities, 4 times replicated in each row.
From each singularity we derived  four $3 \times 3$ templates using a gliding window of this size,
because our aim is to solve our problem with this limited target size.
It would be more natural to use the three singularities directly as $4 \times 4$ templates.
But as we can assemble a singularity by overlaying four $3 \times 3$ templates (the marked windows)
we can get along with the smaller templates of size $3 \times 3$.  
In this way we found the templates J0, J3, J4, J7, K1, and K5. The missing symmetric ones were added
resulting in J0--J7 and K0--K7 as shown in Table \ref{TableTemplatesEven}.

\COMMENT{
\bibitem{2019-pact-domino}
Hoffmann R., Désérable D., Seredyński F. 
(2019).
 A Probabilistic Cellular Automata Rule Forming Domino Patterns. 
In: Malyshkin V. (eds) Parallel Computing Technologies. PaCT 2019. Lecture Notes in Computer Science, vol 11657, pp. 334-344. Springer

\bibitem{Hoffmann-2021} 
Hoffmann, R., Désérable, D. and Seredyński, F.
(2021).
 A cellular automata rule placing a maximal number of dominoes in the square and diamond. 
J Supercomput 77, 9069 – 9087 

\bibitem{Hoffmann-2021-MinimalDominoPact} 
Hoffmann, R., Désérable, D., and Seredyński, F. 
(2021). 
Minimal Covering of the Space by Domino Tiles. In International Conference on Parallel Computing Technologies (pp. 453-465). Springer, Cham.%

\bibitem{Hoffmann-2022-PointPattern} 
Hoffmann, R.,
Forming Point Patterns by a Probabilistic Cellular Automata Rule.
 arXiv preprint arXiv:2202.06656 (2022).

\bibitem{Hoffmann-2002-LoopPattern} 
Hoffmann, R.,
Generating Loop Patterns with a Genetic Algorithm and a Probabilistic Cellular Automata Rule.
Algorithms 16.7 (2023): 352.
}

\section{The CA rule using the templates}

The here designed CA rules 
are tailored to the given problem,
they enhances the construction principles presented already in 
\cite{
2019-pact-domino,
2020-HoffmannSeredynski-SensorPoint,
Hoffmann-2021,
Hoffmann-2021-MinimalDominoPact,
Hoffmann-2022-PointPattern,
Hoffmann-2002-LoopPattern}.
The rule is \textit{probabilistic} and uses \textit{asynchronous} updating. 
A new generation at the next time step $t+1$ is
a sequence of $n^2$ micro time-steps.
The cell state is the tuple

~~~$(a_{ij}, h_{ij})$

where $a_{ij} \in \{0,1\}$ is the \textit{pattern state} and
$h_{ij} \in \{0,1\}$ is the \textit{hit state}.

\COMMENT{
In other words, the center of a local $3\times 3$ configuration has a certain value/utility that 
is given by the SPD game.
Note that here the game is played in asynchronous mode (\textit{dynamically} 
\cite{Wang2022OptimizationGame})
and not in synchronous mode (statically).
The active agent selected at a micro time-step knows all actions already played
in its neighborhood.
}

The rule uses a given set of templates, meaning that 
the templates are pa\-ra\-me\-ters of a general rule. 
It works in the following way.

\begin{itemize}
  \item \textbf{(Select a cell)}
  
  A cell $ij=(i,j)$ is selected at random or sequentially.
  
	\item \textbf{(Compute template hits)}
  
  Each template $\textit{TQ}$ (except its center value $\textit{TQ}_{center}$) is tested for 
  a match at the site $(i,j)$ in the pattern field \textit{A}. 
  More specifically, the outer neighbors of $a_{ij}$ are compared to the spatial corresponding values 
  (also called \textit{pixels}) of each template taken from the given list of templates. 
  
  For each template an associated, temporary hit $h^Q_{ij}$ is stored. It is set to 1 for a hit, otherwise it is set
  to 0. 
  
  Then all  $\textit{h}^Q_{ij}$ are summed up, and if the sum is larger than zero, the hit state $h_{ij}$ is set to one, otherwise to zero. 
  
  
  ~~~if  $(\sum_{\forall Q} h^Q_{ij} )>0$ then $h_{ij} := 1$ else  $h_{ij} := 0$ 
  
  \item \textbf{(Adjust or Inject Noise)}
  
    \begin{itemize}
      \item \textbf{(If there is a hit: Adjust)}
      
      If  $h_{ij}=1$ then there exists a template hit $h^Q_{ij} =1$.
      
            \begin{itemize}
              \item           
               If there is only one template hit, the pattern state is set to the center value of the template.
          
               ~~~if $h^Q_{ij} =1$ then $a_{ij}:=\textit{TQ}_{center}$.
            
               This operation is not necessary if
              the  pattern state  $a_{ij}$ is already equal to the correct value  $\textit{TQ}_{center}$,
               which means that the complete template was found in the pattern.
              
              \item
              If there are several template hits, one of them is selected at random for adjustment. 
            
            \end{itemize}

      \item \textbf{(If there is no hit: Inject Noise)}
      
			In order to drive the evolution to optimal patterns, noise needs to be injected
			into the CA.
			This is performed by changing the pattern state $a_{ij}$ with some randomness.
			By default the current state of a cell shall be unchanged. 
			In general, for each possible state transition $A \rightarrow B, A\neq B$ 
			a specific probability needs to be defined, that may further depend
			on an additional condition. In our case we have only the two states
			zero and one, and so we need to define the state transitions 
			$0 \rightarrow 1$, and $1 \rightarrow 0$:

        ~~~if $a_{ij}=0$ then  $a_{ij}:=1$ with probability $\pi_{01}$
        
        ~~~if $a_{ij}=1$ then  $a_{ij}:=0$ with probability $\pi_{10}$	.
			
			In our simulations, the best results were obtained with the settings
			 $\pi_{01}=0.04$ and $\pi_{10}=1.0$.
			So there is an unconditional change $1\rightarrow 0$, and
			a state change $0\rightarrow 1$ with a low probability of $0.04$.
			In terms of the PD game, the state 1 means ``Defect'', and 0 means ``Cooperation''. 
      
    \end{itemize}

\end{itemize}

Depending on the used subset of all the 52 templates (Table~\ref{TableTemplatesEven}), 
we defined the following specialized rules

\begin{itemize}
  \item \textbf{Rule 8:} 
  This rule uses only the templates T0--T7, 
  it is dedicated to work for patterns of even size $n$ only.  
  
	\item \textbf{Rule 36:}    
  This rule uses the templates T0--T35.
  It  forms \textit{transient} optimal or near-optimal patterns of odd size. 
  The templates describing singularities are not used. 

	\item\textbf{Rule 52:}  
  This rule uses the the full set of templates T0--T51.
  It forms \textit{stable} optimal or near-optimal patterns of odd size.   
  
\end{itemize}

The CA rule is interpreted as follows.

\footnotesize
\begin{itemize}
	\item \textbf{Initialize}

	\begin{itemize}
		\item []
			\textbf{for all} $(i,j)$ \textbf{do}
			
			~~~$a_{ij}\leftarrow1$ with probability 1/4, else $a_{ij}\leftarrow0$
			
		  ~~~$h_{ij}\leftarrow0$~~~---\textit{The hits are reset}	
			
				
			\textbf{end for}
		
		\item[]
			\textit{Evaluate global objective }$^{(*)}$
	\end{itemize}
	
	\item
	\textbf{for} $t\leftarrow1$ \textbf{to} $t_{limit}$ \textbf{do}
	
	\begin{itemize}
		\item[] 
		
		\textbf{repeat} $n^2$ \textbf{times}~~~--- \textit{compute a new generation, $n^2$ micro time-steps}
    
    ~~~Select a cell $(i,j)$ at random
    
    ~~~$\forall Q: h^Q_{ij} \leftarrow 0/1$~~~--- \textit{The results of testing all templates TQ }
		 
		~~~$h_{ij} \leftarrow$ Number of template hits
		
		~~~$a_{ij} \leftarrow$ Rule($h_{ij}$, Center of template TQ with hit $h^Q_{ij}=1$)~--- \textit{Adjust or Noise}			
		
		\textbf{end repeat}					
						
		\item[] 	
     $^{(*)} $\textit{Evaluate global objective}
        
     ~~~$\forall (i,j): \textit{compute}~ u_{ij}\leftarrow(a_{ij}^{+9})/9$     
    
		~~~\textit{STP(A)} and $W(A)$ depending on $\sum u_{ij}$ are computed.
	\end{itemize}
		
\textbf{end for}		
	
\end{itemize}
\normalsize

Recall that $n^2$ micro time-steps are performed  in asynchronous updating mode per generation.
The utilities $u_{ij}$, \textit{STP(A)} and $W(A)$ are only computed for evaluation and statistics, they do not
influence the pattern evolution. 
The pattern states $a_{ij}$ and the hit states $h_{ij}$
are stored between the generations.
During the computation of a new generation not all cells are updated,
roughly 1/3 never, 1/3 once, and 1/3 multiple times, according to the binomial distribution. 
Therefore the states ($h_{ij}^k$ and  $u_{ij}^k$) of a neighbor $k$  
can be depreciated because changes of their neighborhoods between their last update and the current 
updating may not be reflected. 
This effect can be seen as an additional source of noise. 
Fortunately, in our simulations, this  effect did not disturb the overall convergence to 
a stable pattern.
When the pattern becomes stable, this perturbing effect vanishes.

\section{Patterns evolved by the CA rule}
\subsection{Even pattern sizes}

\begin{figure}[H] 
\centering
(a)

\includegraphics[width=\textwidth]{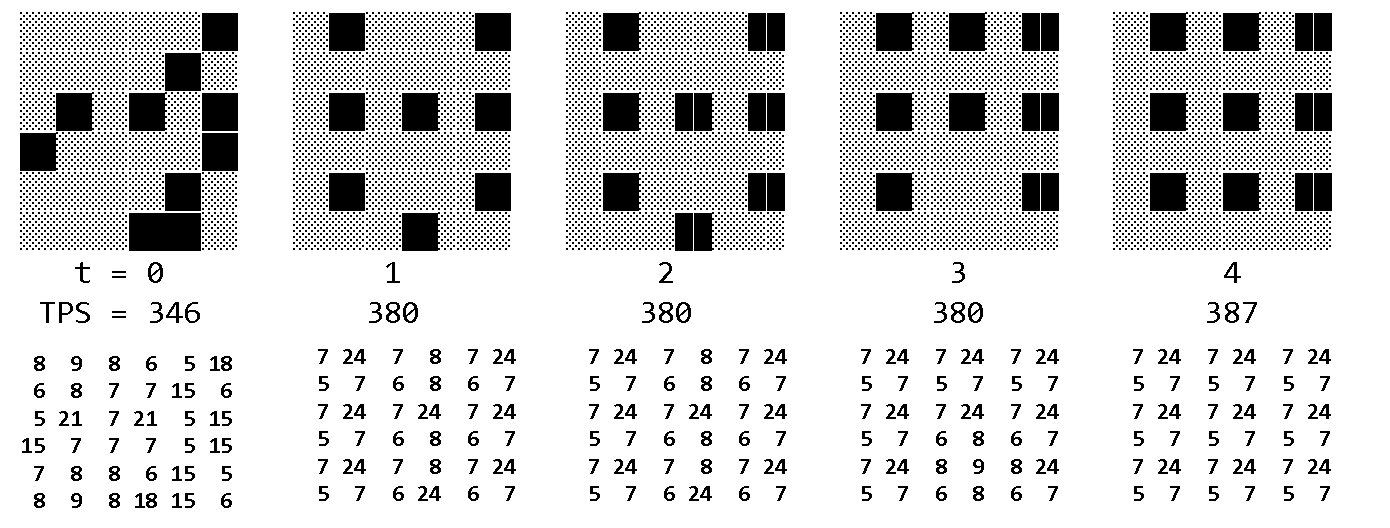}
	
(b)

\includegraphics[width=\textwidth]{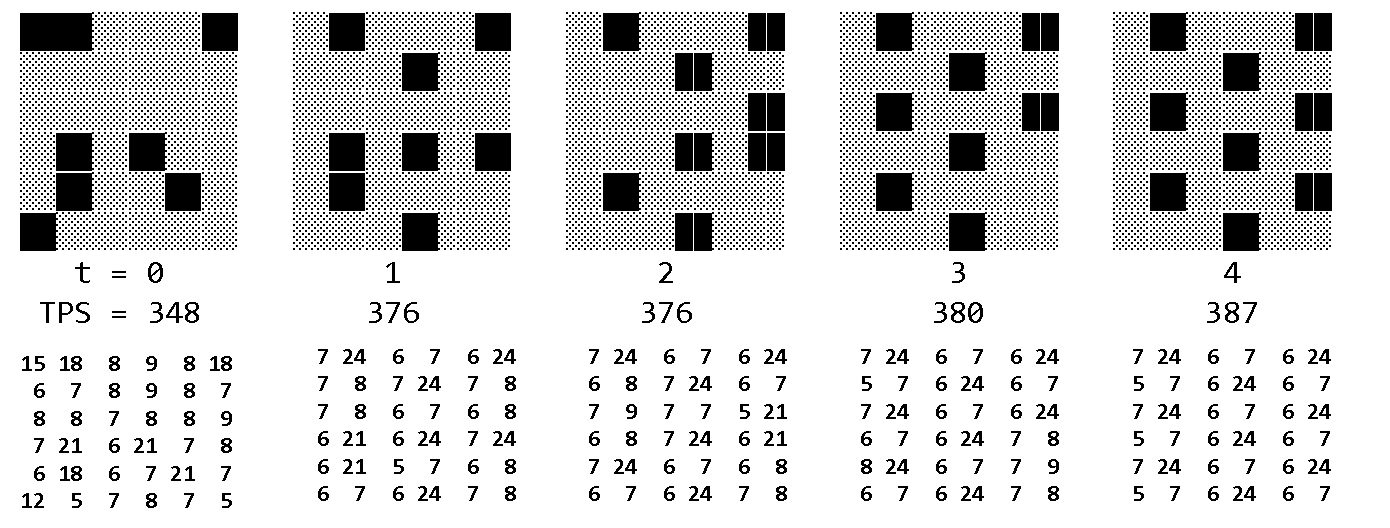}	
\caption{
 The evolution of two  
optimal $6 \times 6$ patterns (a) and (b) using rule 8.  
Wealth at $t=4$ is $W= 1.1944$, and $TPS=387$.
The total payoffs are shown below the patterns. 
}
\label{6x6snapshot}
\end{figure}

\textbf{Application of Rule 8.}
At first we want to use Rule 8 that was designed to evolve patterns of even size $n$.  
The initial pattern is generated randomly. 
The patterns cell states are set to 1 with the probability of 1/4.
This probability was used because  optimal pattern have 
a density close to 1/4 and this setting may help to speed-up the evolution. 
Note that the real initial rate of ones can deviate from $n^2/4$
because the distribution is binomial. 
The Rule 8 drives the evolution to a stable optimal pattern. 
Fig.~\ref{6x6snapshot}
shows two sample evolutions with 4 generations only, from $t=0$ to $t=4$.
The payoffs for every cell are  also given. 
In an optimal pattern the 1-cells receive a payoff of $N_{zero} \times T =24$,
and the 0-cells $(1+N_{zero})\times R=5,6,$ or $7$,
where $N_{zero}$ denotes the number of zeroes in the outer Moore-neighborhood
with 8 neighbors.

100 runs (experiments) were performed in order to compute the average time (the number of time-steps)
to reach a stable optimum. The simulation time limit was set large enough to reach the optimum
in every run. 

The measured values are:

~~~$t_{avrg}^{opt}(n=6) = 34~(3_{min}-200_{max})$.

~~~$t_{avrg}^{opt}(n=8) = 198~(5_{min}-733_{max})$.

~~~$t_{avrg}^{opt}(n=10) = 1421~(31_{min}-4998_{max})$.

Note that this values are not absolutely certain because of the statistical variations that 
have to be taken into account. 
Nevertheless we can see that the time is increasing very rapidly.
What can we do against it?
If we are searching only for one or a few optimal solution we can 
make a number of runs $N_{run}$ with a limited number of generations $t_{limit}$ and then select the 
optimal or best ones. 

For performance evaluation we can compute the rate 
$N^{opt}_{found} \leq N_{run}$ of found optimal solutions within $N_{run}$ runs for a limited number $t_{limit}$
of generations. 

Another way to evaluate the performance is to compute for a given Rule 8/36/52, a given size $n$, 
and a given time limit  $t_{limit}$  

(i) $W^{max}_{run}$,  the maximal wealth 
 found in each \textit{run}, 
found first at $t=t_{max}$~, and 
  
 (ii)  $W^{max}_{avrg}= 1/N_{run}\sum_{\forall run}W^{max}_{run}$, the  maximal obtained wealth
averaged over all runs. 

(iii) $W^{max}_{max}$, the best pattern found during all runs.  

\begin{itemize}
	\item 
      \textit{Example}: For $n=10$ we run $N_{run}=100$ evolutions (runs) with different $t_{limit}$. 
      For $t_{limit} =(50, 100, 200, 500, 1000, 5000)$ we found 

      ~~~$N^{opt}_{found}=(4,6,13,30,55,100)$ optimal solutions.

      The ratio $N^{opt}_{found}/N_{run}=\textit{expect}(N^{opt})$ approximates 
      the expectation of finding an optimal solution, which is in our example

      ~~~$\textit{expect}(N^{opt}, t_{limit}) = (4/100, 6/100, 13/100, 30/100, 55/100, 100/100)$ 

      ~~~$= (4\%, 6\%, 13\%, 30\%, 55\%, 100\%)$.

      The achieved maximal wealth, averaged over all runs, and for the given time limits are

       ~~~$W^{max}_{avrg}= (1.1709, 1.1745, 1.1814, 1.1870, 1.1905, 1.1944)$.

      For this example, during $t_{limit}=50$ iterations, we can expect with a probability of $4\%$ an optimal pattern with
      wealth 1.1944, and a pattern with wealth of 1.1709 on average. 
      This means that we can find very good performing patterns within a short time. 
      Recall that the baseline pattern has the wealth of 1,
      it is completely filled with zeroes (cooperators in terms of the PD game).

\end{itemize}

\textbf{Application of Rule 52.}
We want to test whether the rules designed for patterns of odd size can also work. 
First we test Rule 52.
We use the same example with $n=10$, $N_{run}=200$.
The result is:
\begin{itemize}
	\item 
  All patterns reach quickly a \textit{stable near-optimal} configuration.
  The  time needed is $t_{avrg}= 10 ~(2 -74)$.
  \item
  The best pattern found has $W^{max}_{max}=1.1922$. 
  It is close to the optimum.
  The average of the run's maxima is $W^{max}_{avrg}=1.1841$. 
  \item
  The measured wealth distribution $((W, rate))$ was
  
    \footnotesize
    \begin{verbatim}
(1.1755, 1),(1.1766, 1),(1.1777, 2), (1.1788, 3), (1.1800, 10),(1.1811, 10), 
(1.1822, 8),(1.1833, 6),(1.1844, 17),(1.1855, 15),(1.1866, 9), (1.1877, 9), 
(1.1888, 2),(1.1900, 3),(1.1911, 2), (1.1922, 2)   .
    \end{verbatim}
    \normalsize   

\end{itemize}

\begin{figure}[H] 
\centering
\includegraphics[width=9cm]{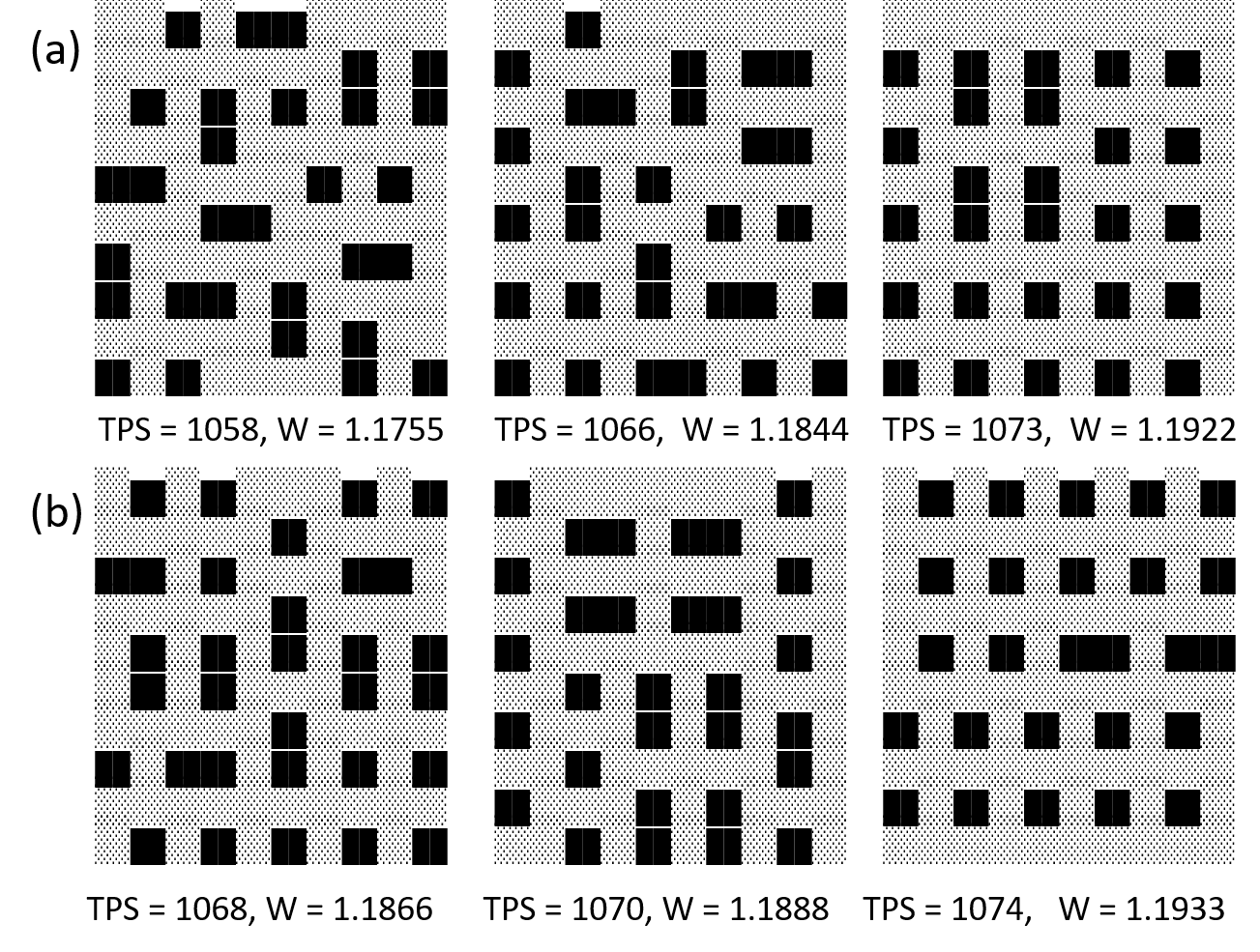}	
\caption{
(a) Some near-optimal $10 \times 10$ patterns found with (a) Rule 52, and 
(b) with Rule 36.
}
\label{10x10nonoptR52R36}
\end{figure}

We found a variety of good patterns close to the optimum which we did not reach,
although it can be reached in principle. 
Some of these patterns are shown in Fig.~\ref{10x10nonoptR52R36}(a).
		
\textbf{Application of Rule 36.}
We want also test Rule 36,
using the same example with $n=10$, $N_{run}=250$.
The result is:
\begin{itemize}
	\item 
    All evolved patterns were \textit{stable} and \textit{near-optimal}.
    The  time needed is $t_{avrg}= 54 ~(4 -236)$.
    \item
    The best pattern found has $W^{max}_{max}=1.1933$. 
    It is very near to the optimum, probably it is the nearest.
    The average of the run's maxima is $W^{max}_{avrg}=1.1897$. 
    \item
    The measured wealth distribution $((W, rate))$ was
      
    \footnotesize
    \begin{verbatim}
(1.1866, 3),  (1.1877, 13), (1.1888, 30), (1.1900, 23),
(1.1911, 19), (1.1922, 8),  (1.1933, 4).
    \end{verbatim}
    \normalsize   

\end{itemize}

We found a variety of stable near-optimal patterns, significantly better than
with Rule 52. The reason is that we do not allow singularities, therefore these patterns 
are more dense. Compared to Rule 52, the evolution lasts longer (54:10). 
Some of these patterns are shown in Fig.~\ref{10x10nonoptR52R36}(b).

\subsection{Odd pattern sizes}
\label{Odd pattern sizes}

Rule 8, designed for even pattern sizes, does not evolve stable patterns for odd sizes, it tries to place points surrounded by zeroes of size $3 \times 3$, but these points are transients only.
Therefore Rule 36 and 52 were designed which can evolve optimal patterns. 
Rule 52 forms \textit{stable} optimal and near-optimal patterns, 
and Rule 36 \textit{transient} optimal and near-optimal patterns.
In the following some evolved patterns are presented which can be found with both rules.  
The CA rules can evolve the same patterns that we found by the Genetic Algorithm.

\begin{figure}[H] 
\centering
\includegraphics[width=9cm]{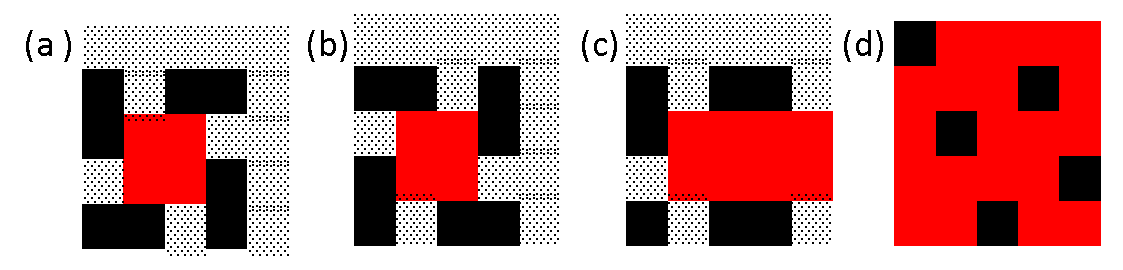}	
\caption{
$5 \times 5$ patterns evolved by the Rules 52 and 36. 
(a, b) are optimal patterns with $W=1.1777$,
(c, d) are near-optimal ones. 
Wealth is $W= 1.1644$ for (c), and 1.1555  for (d). 
$2\times2$ zero-blocks are marked in red.
}
\label{CApattern5x5}
\end{figure}

\textbf{Size $5\times 5$.}
There is only one optimal pattern, shown in Fig.~\ref{CApattern5x5}(a, b).
(a) and (b) are symmetric under reflection, so they are counted only once. 
We can see the red marked $2\times 2$ square of zeroes which marks a singularity.
Two non-optimal patterns are shown in Fig.~\ref{CApattern5x5}(c, d).

\begin{figure}[H] 
\centering
\includegraphics[width=9cm]{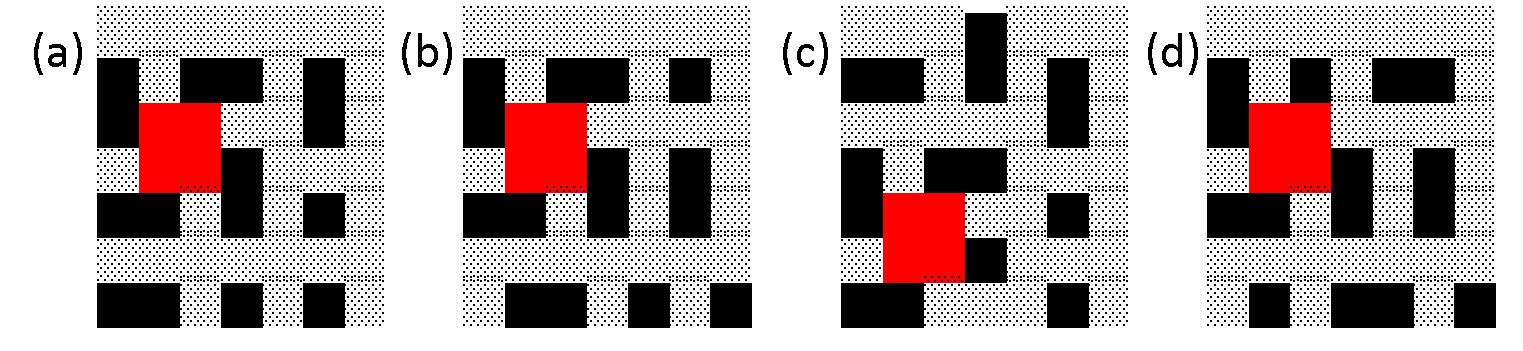}	
\caption{
Some optimal $7\times 7$ patterns. Wealth is $W= 1.1836$  and $TPS=522$.
 $2 \times 2$ zero-squares are marked in red.
}
\label{optimal7x7}
\end{figure}

\textbf{Size $7\times 7$.}
Four optimal patterns are depicted in Fig.~\ref{optimal7x7}.
The $4 \times 4$ singularities in (a, b) are similar
but different to (c, d).

\begin{figure}[H] 
\centering
\includegraphics[width=9cm]{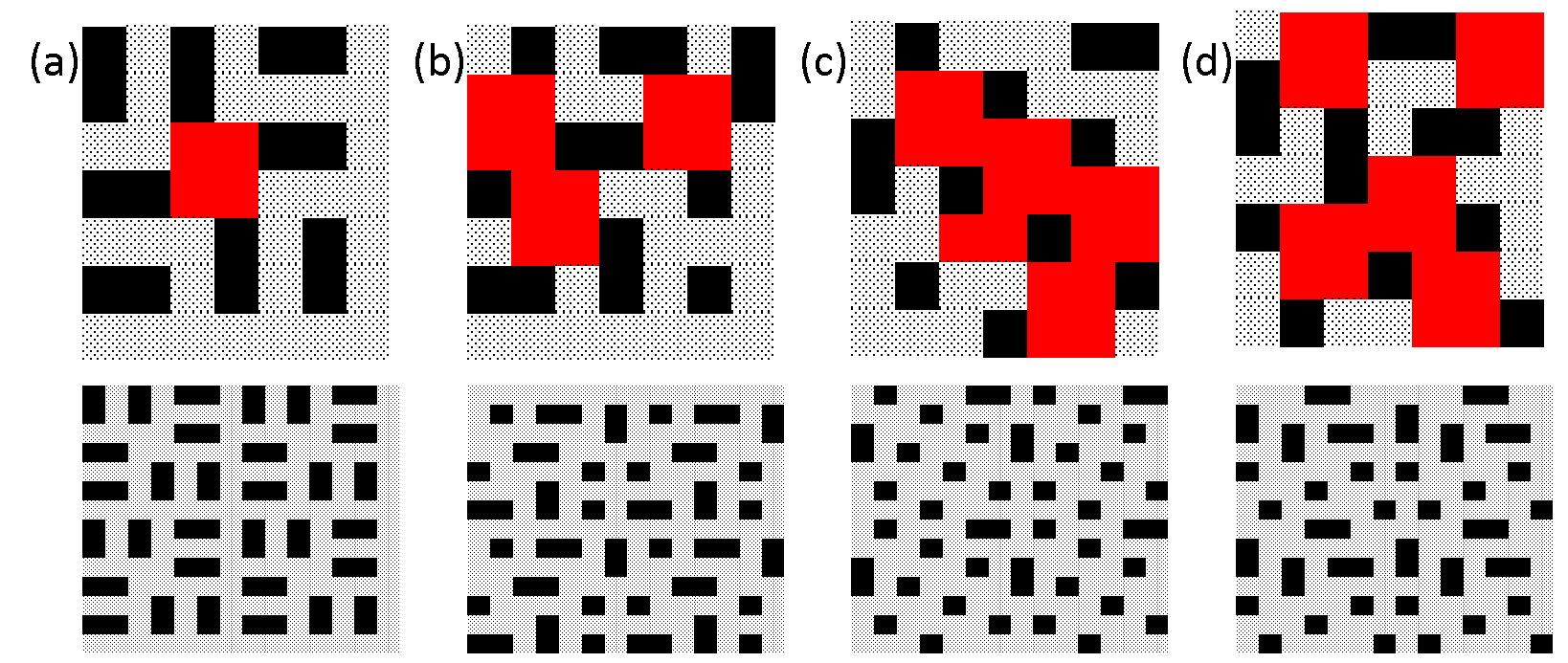}	
\caption{
Some near-optimal $7 \times 7$ patterns.
Wealth of $(a, b, c, d)$ is $(1.1814, 1.1768, 1.1723, 1.1700)$
and \textit{TPS}$ = (521, 519, 517, 516)$. 
The corresponding quad representation is shown in the bottom row. 
Zero-blocks are marked in red.
}
\label{nonopt7x7}
\end{figure}

Near-optimal patterns (Fig.~\ref{nonopt7x7}) 
can also be of interest, for instance from the artistic point of view.
Pattern (a) contains dominoes only, has only one singularity and is highly symmetric,
a rare case.
This case shows that patterns with one singularity need not necessarily to be optimal,
though we observed \textit{exactly one} singularity in \textit{optimal} patterns of odd size.

\begin{figure}[H] 
\centering
\includegraphics[width=9cm]{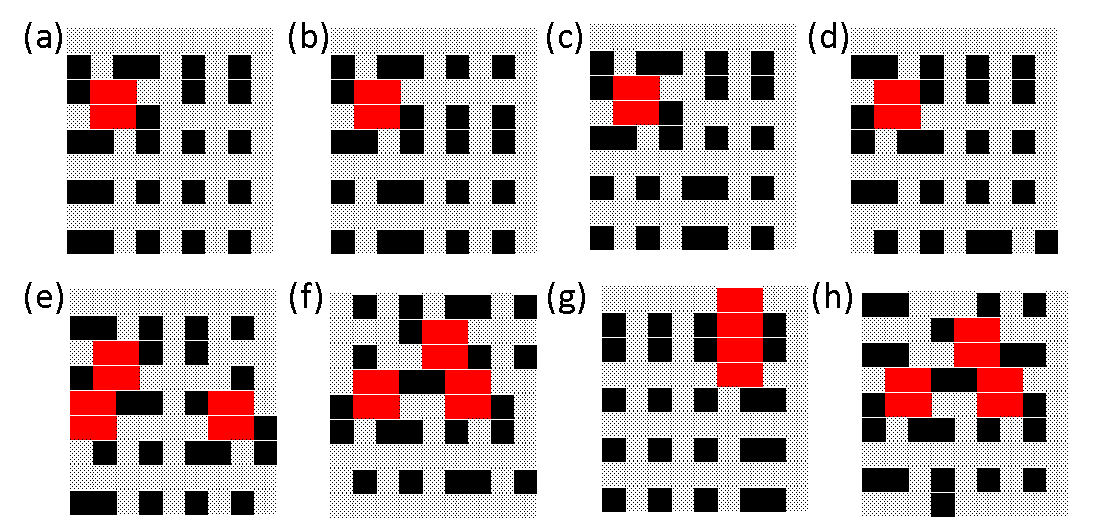}	
\caption{
$9 \times 9$ patterns evolved by CA. 
(a)--(d) are optimal patterns with $W=1.16615$,
(e)--(f) are near-optimal ones. 
(e)--(g) have a wealth $W= 1.1824$, and (h) a wealth of  1.180.
Zero-blocks are marked in red.
}
\label{CApattern9x9}
\end{figure}

\textbf{Size $9\times 9$.}
Fig.~\ref{CApattern9x9}(a--d) shows optimal patterns,
and (e--f) near-optimal ones. 
The optimal patterns contain only one singularity. 

\begin{figure}[H] 
\centering
\includegraphics[width=8cm]{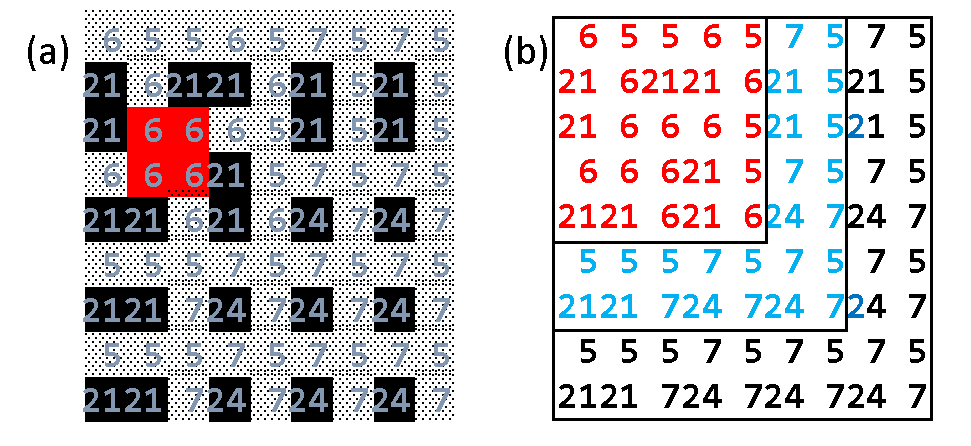}	
\caption{
(a) $9 \times 9$ optimal pattern with payoffs. 
(b) This specific pattern can recursively be constructed from 
the $5\times 5$ pattern (with payoffs in red) by adding 
two rows and two columns (blue marked payoffs) to get
the $7\times 7$ pattern, and then adding again two rows and columns 
(black marked payoffs) to get
the $9\times 9$ pattern.
}
\label{CApattern9x9WithPayoff}
\end{figure}

\textbf{Constructing optimal odd patterns.}
Can optimal patterns be constructed? Yes, see Fig.~\ref{CApattern9x9WithPayoff}.
We start with the $5\times5$ pattern, and then build a $7\times7$ pattern.
(i) A row  $0^5$   and a column $011(01)^1$ is added.
(ii) A row $110(10)^1$  and a column $0^5$ is added.
(iii) The missing square $^{00}_{10}$ is added. 
In a similar way the $9\times9$ pattern can be constructed, and so on. 
This construction principle is simple and can easily be generalized. 

Let $m=(n-5)/2$, $n \geq 5$, 
defining the relation $(n,m)= (5,0), (7,1), (9,2) ...$.
Let $n \times n$ be the size of the next larger pattern to be constructed.
Then the construction principle is
(i) add row $110(10)^m$  and column $0^{n-2}$,
(ii) add row $110(10)^m$  and column $0^{n-2}$, 
(iii) add the missing square $^{00}_{10}$.

\textbf{A general formula for the total play sum TPS.}
We want to know the maximal TPS for any odd size $n$.
Recall that TPS (the fitness in the Genetic Algorithm) is 

~~~$\textit{TPS} = W\times n^2 \times (K=9)$.
 
To provide the TPS of our PD game we regard the payoffs
shown in Fig.~\ref{CApattern9x9WithPayoff}. 
We count 265 for the basic $5\times5$ area,
64 for a $2 \times 3$ rectangles at the borders with the values $^{~5~~5~~5}_{21~21~~7}$,
and 43 for squares with $^{~7~~5}_{24~~7}$. In total we get

~~~$\textit{TPS} = 265 + 128m + 43(m(m+2))$.

This formula gives us the exact value of an optimal solution. 
For very large $m=(n-5)/2$, \textit{TPS} approximates $43m^2$.
Dividing TPS by $Kn^2$ and computing the limit for large $n$
gives the limit wealth $W^{limit}=43/36= 1.19444$. This means
that for odd pattern the TPS is always smaller than this limit
which is the optimal value for the even case. 

By visual inspection we observe that such an optimal odd pattern 
contains $n-1$ dominoes at two neighboring border sides, 
and is  elsewhere (in the remaining square of the whole pattern field) filled with  points
$\substack{000
         \\010
         \\000}$.          
        
\textbf{Performance of  Rule 52.}     
We simulated the case $n=9$, $N_{run}=100$, $t_{limit}=100$.
The result is:
\begin{itemize}
	\item 
    All patterns reach quickly a \textit{stable} optimal or near-optimal configuration.
    The  time needed was $t_{avrg}= 11 ~(2 -49)$.
    \item
    The best patterns found have $W^{max}_{max}=1.1865$, the optimum.
    The average of the run's maxima was $W^{max}_{avrg}=1.1834 ~(99.74\%)$. 
    \item
    The measured wealth distribution $((W, rate))$ was  
      
    \footnotesize
    \begin{verbatim}
(1.1769, 5),  (1.1783, 5), (1.1796, 2),  (1.1810, 17), 
(1.1824, 22), (1.1838, 1), (1.1851, 14), (1.1865, 34)! .
    \end{verbatim}
    \normalsize  

\end{itemize}
34 of 100 of the found pattern are optimal, the remaining a near-optimal.

Now we want to know how the Rule 52 performs for a larger field size. 
Another simulation with $n=27$, $N_{run}=100, t_{limit}=100$ was conducted.
The number of cell states to be computed is now 9 times higher
compared to $n=9$. The results are:

\begin{itemize}
  \item
  All patterns reach quickly a stable configuration.
  The  time needed was $t_{avrg}= 31 ~(9-68)$.  

	\item 
  No optimal pattern with $W=1.1922$, $\textit{TPS}=7822$ was found. 
  
  \item
   The average of the run's maxima is $W^{max}_{avrg}=1.1833 ~(99.25\%)$,
  $W^{max}$ ranges between 1.1806 and 1.1867 (\textit{TPS} between 7748 and 7786).   
  
\end{itemize}

\textbf{Performance of  Rule 36.}   
We simulated the case $n=9$, $N_{run}=100$, $t_{limit}=100$.
The result is:

\begin{itemize}
	\item 
  All evolved patterns are \textit{transient} but of very high quality.
  The average time needed to find the best in every run was $t_{avrg}= 26 ~(2-88)$.
  \item
  The optimal pattern with $W^{max}_{max}=1.1865$ was found 96 times. 
  The average of the run's maxima was $W^{max}_{avrg}=1.1864$. 
  \item
  The wealth distribution $((W, rate))$ was   
  \footnotesize
  \texttt{(1.1851, 4),  (1.1865, 96).}
  \normalsize 

\end{itemize}

Comparing Rule 36 to Rule 52, we found (96:34) optimal patterns, though they are transients. 
The average time to find the solutions was (26:11).

A second simulation case  was conducted with $t_{limit}=50$ instead of 100.
Optimal patterns were found 80 times, and 
$t_{avrg}= 20 ~(3-50)$, $W^{max}_{avrg}=1.1858$.
Then a third  simulation experiment  was conducted with $t_{limit}=10$.
Optimal patterns were found 32 times, and 
$t_{avrg}= 5 ~(3-10)$, $W^{max}_{avrg}=1.1825$.
The results demonstrate that Rule 36 is able to outperform Rule 52 although the 
evolved optimal patterns are not stable.

Now we want to know how the Rule 36 performs for a larger field size. 
Another simulation with $n=27$, $N_{run}=100, t_{limit}=100$ was conducted.
The results are:

\begin{itemize}
  \item
  The evolved patterns are not stable.
  The  time needed for the best patterns within the time limit 
  was $t_{avrg}= 76 ~(37-100)$.  

	\item 
  No optimal pattern with $W=1.1922$, $\textit{TPS}=7822$ was found. 
  
  \item
   The average of the run's maxima was $W^{max}_{avrg}=1.1880 ~(99.65\%)$,
  $W^{max}$ ranges between 1.1870 and 1.1899 (\textit{TPS} between 7788 and 7807). 
  
\end{itemize}

\begin{figure}[H] 
\centering
\includegraphics[width=4cm]{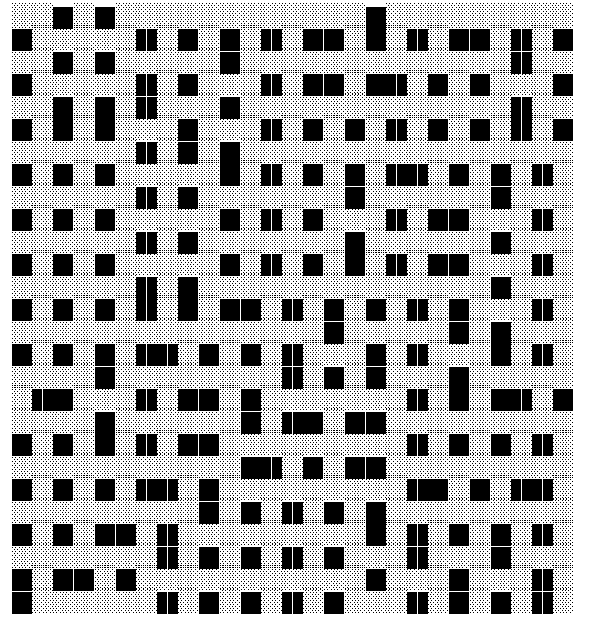}	
\caption{
A near-optimal pattern of size $27 \times 27$ 
with  $W=1.1899$, $\textit{TPS}=7807$, 
with 98 points, 56 dominoes,
519 zero-states, and 210 one-states
}
\label{27x27}
\end{figure}

One of the found near-optimal patterns is shown in Fig.~\ref{27x27}.
There is no singularity in this pattern.

\begin{figure}[H] 
\centering
\includegraphics[width=\textwidth]{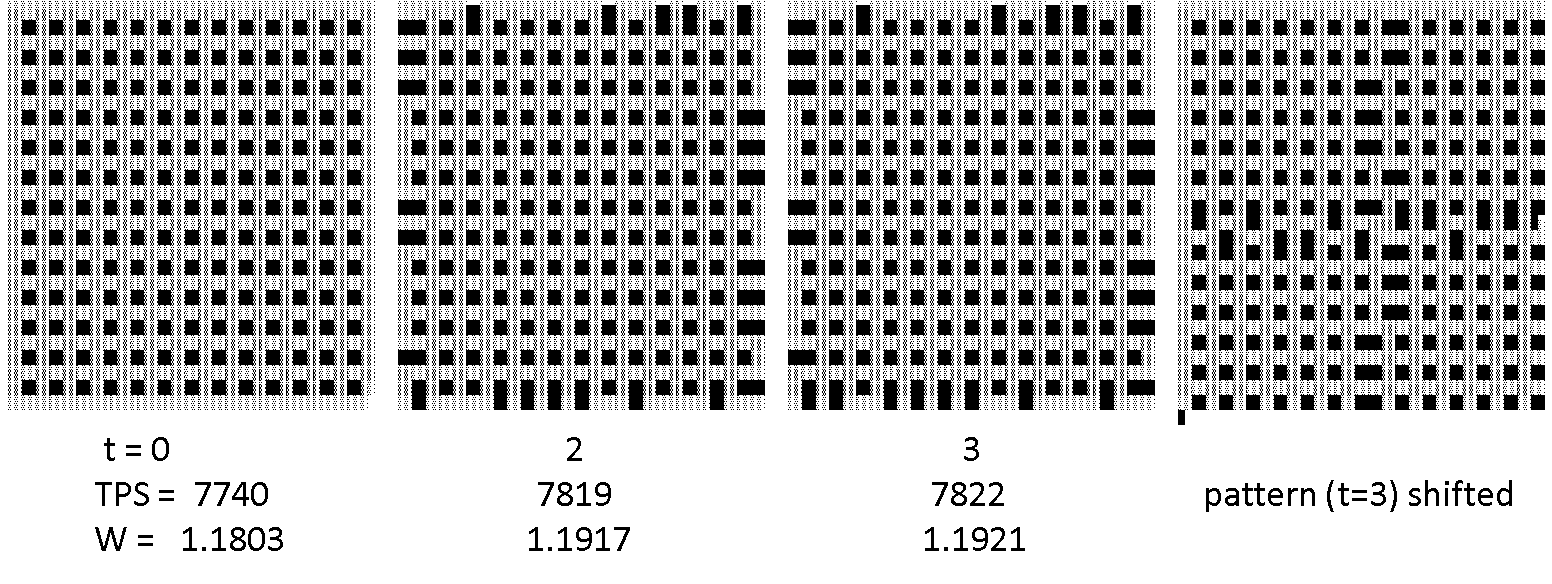}	
\caption{
A fast evolution of an optimal  $27 \times 27$ pattern.
with  $W=1.11921$, $\textit{TPS}=7821$. 
The initial pattern at $(t=0)$ was filled with points. 
}
\label{27x27initpointquad}
\end{figure}

Another experiment was conducted in order to find easily optimal patterns.
The initial configuration was totally filled with points
with the mandatory extra spaces at the borders (because $n$ is odd).
From this configuration  optimal configurations (Fig.~\ref{27x27initpointquad}) 
 evolve rapidly within 
approx. 3--50 generations using Rule 36 or Rule 52. 
The singularity in 
Fig.~\ref{27x27initpointquad}
for the final generation $(t=3)$
can better be detected after having shifted it.

\section{Conclusion}

A CA rule was developed in a methodical way that can evolves optimal patterns
with respect to a global fitness function.
The fitness function was declared as the sum of local utilities computed by the cells/agents.
A utility is a local value function over all states in the neighborhood.
Here the utility was defined as the payoff that a player would receive in
a Spatial Prisoner's Dilemma Game with a special parameter setting. 
The used method performs in three steps:
(1) optimal ``master'' patterns are generated by a genetic algorithm,
(2) templates (local matching patterns) are extracted from the master patterns, and
(3) the templates are inserted into a general probabilistic CA rule.

The CA rule for even grid sizes needs only 8 templates of size $3\times 3$,
whereas it needs 36 or 52 templates to cope with grids of odd sizes.
The rules can evolve optimal or near-optimal patterns.
Depending on the selected rule, the patterns are stable or transient.
Optimal patterns of even size contain only points (a one surrounded by zeroes).
Optimal patterns of odd size are a mixture of points and dominoes,
and they contain exactly one singularity, a framed square of four zeroes. 

In \textit{future work} the following topics could be addressed.
\vspace{-7pt}
\begin{itemize}
	\item 
  How do other parameters \textit{T} and \textit{P} impact the resulting patterns?
  \item
  Another utility function not necessarily related to a game could be used.
  \item
  Where are the limits of this approach?  
  \item
  What is the minimal necessary size of the templates, and can larger sizes speed-up the
  evolution?
  \item
  Templates could be defined by local utility configurations instead of cell state configurations.
  Is there an advantage?
  
  \item
  The design process could be further automated.
\end{itemize}


\end{document}